\documentclass[a4paper,11pt]{article}
  \pdfoutput=1 

\usepackage{jheppub} 

\usepackage[T1]{fontenc} 

\usepackage[utf8]{inputenc}
\DeclareUnicodeCharacter{27E8}{\ensuremath{\langle}}
\DeclareUnicodeCharacter{27E9}{\ensuremath{\rangle}}
\DeclareUnicodeCharacter{0394}{\ensuremath{\Delta}}
\usepackage{amsmath, amsthm, mathtools,amssymb}
\usepackage{hyperref}
\usepackage{multicol,multirow}
\usepackage[sort&compress,numbers]{natbib}
\usepackage{subfigure}

\newcommand\fverb{\setbox\fverbbox=\hbox\bgroup\verb}
\newcommand\fverbdo{\egroup\medskip\noindent%
			\fbox{\unhbox\fverbbox}\ }
\newcommand\fverbit{\egroup\item[\fbox{\unhbox\fverbbox}]}
\newbox\fverbbox

\def\verytiny{\kern.08em}

\def\bentarrow{\:\raisebox{1.3ex}{\rlap{$\vert$}}\!\rightarrow}
\def\bothdk#1#2#3#4#5{
\begin{array}{r c l}
#1 & \rightarrow & #2#3 \\
 & & \:\raisebox{1.3ex}{\rlap{$|$}}\raisebox{-0.5ex}{$|$}%
\phantom{#2}\!\bentarrow #4 \\
 & & \bentarrow #5
\end{array}
}
\def\ib{{\; \bar\imath}}
\def\ctw{\cos\theta_W}
\def\stw{\sin\theta_W}
\def\sstw{\sin^2\theta_W}
\def\cstw{\cos^2\theta_W}
\def\DeltaThree{\Delta_3}

\def\sep{\times}

\def\lbar{\bar{\ell}}
\def\qbar{\bar{q}}
\def\beq{\begin{equation}}
\def\eeq{\end{equation}}

\def\I33m{\mathrm{I}_3^{3{\mathrm m}}}
\def\nn{\nonumber}

\def\be{\begin{equation}}
\def\ee{\end{equation}}
\def\beqn{\begin{eqnarray}}
\def\eeqn{\end{eqnarray}}
\def\bea{\begin{eqnarray}}
\def\eea{\end{eqnarray}}

\def\spa#1.#2{\left\langle#1#2\right\rangle}
\def\spab#1.#2.#3{\langle#1|\Gamma_{#2}|#3]}
\def\spaba#1.#2.#3.#4{\left\langle{#1}|\Gamma_{#2|#3}|#4\right\rangle}
\def\tspaba#1.#2.#3.#4{\langle{#1}|\tilde{\Gamma}_{#2|#3}|#4\rangle}
\def\spabab#1.#2.#3.#4.#5{\left\langle#1|\Gamma_{#2|#3|#4}|#5\right]}
\def\spb#1.#2{\left[#1#2\right]}
\def\spba#1.#2.#3{[#1|\Gamma_{#2}|#3\rangle}
\def\spbab#1.#2.#3.#4{[#1|\Gamma_{#2|#3}|#4]}
\def\tspbab#1.#2.#3.#4{[#1|\tilde{\Gamma}_{#2|#3}|#4]}
\def\spbaba#1.#2.#3.#4.#5{[#1|\Gamma_{#2|#3|#4}|#5\rangle}
\def\tspbaba#1.#2.#3.#4.#5{[#1|\tilde{\Gamma}_{#2|#3|#4}|#5\rangle}
\def\lor#1.#2{\left(#1#2\right)}
\def\sand#1.#2.#3{%
\left\langle\smash{#1}{\vphantom1}^{-}\right|{#2}%
\left|\smash{#3}{\vphantom1}^{-}\right\rangle}
\def\sandp#1.#2.#3{%
\left\langle\smash{#1}{\vphantom1}^{-}\right|{#2}%
\left|\smash{#3}{\vphantom1}^{+}\right\rangle}
\def\sandpp#1.#2.#3{%
\left\langle\smash{#1}{\vphantom1}^{+}\right|{#2}%
\left|\smash{#3}{\vphantom1}^{+}\right\rangle}
\def\sandpm#1.#2.#3{%
\left\langle\smash{#1}{\vphantom1}^{+}\right|{#2}%
\left|\smash{#3}{\vphantom1}^{-}\right\rangle}
\def\sandmp#1.#2.#3{%
\left\langle\smash{#1}{\vphantom1}^{-}\right|{#2}%
\left|\smash{#3}{\vphantom1}^{+}\right\rangle}
\def\spaaold#1.#2.#3{\langle#1|#2|#3\rangle}
\def\spbbold#1.#2.#3{[#1|#2|#3]}
\def\spaa#1.#2.#3.#4{\left\langle#1|#2|#3|#4\right\rangle}
\def\spbb#1.#2.#3.#4{\left[#1|#2|#3|#4\right]}
\def\spaxa#1.#2.#3.#4{\langle#1|#2|#3|#4\rangle}
\def\spbxb#1.#2.#3.#4{[#1|#2|#3|#4]}

\def\tree{\mathrm{tree}}
\def\oneloop{\mathrm{1-loop}}

\def\prp34{\mathcal{P}_{34}}
\def\prpp345{\mathcal{P}_{345}}

\usepackage{slashed}
\allowdisplaybreaks

\preprint{\begin{minipage}[t]{8cm}\begin{flushright}FERMILAB-PUB-22-231-T,\\FR-PHENO-2022-06,\\ IPPP/22/17\end{flushright}\end{minipage}}
\title{\boldmath Vector boson pair production at one loop:
  analytic results for the process $q \bar{q} \ell \bar\ell \ell^\prime \bar{\ell}^\prime g$}

\author[a]{John M. Campbell,}
\emailAdd{johnmc@fnal.gov}

\author[b]{Giuseppe De Laurentis,}
\emailAdd{giuseppe.de.laurentis@physik.uni-freiburg.de}

\author[c]{R. Keith Ellis}
\emailAdd{keith.ellis@durham.ac.uk}

\affiliation[a]{Fermilab, PO Box 500, Batavia IL 60510-5011, USA}
\affiliation[b]{Physikalisches Institut, Albert-Ludwigs-Universit{\"a}t, D-79104 Freiburg, Germany}
\affiliation[c]{Institute for Particle Physics Phenomenology, Durham University, Durham, DH1 3LE, UK}
\date{\today}

\abstract{We present compact analytic results for the one-loop
  amplitude for the process $0 \rightarrow q \bar{q} \ell \bar\ell
  \ell^\prime \bar{\ell}^\prime g$, relevant for both the production
  of a pair of $Z$ and $W$-bosons in association with a jet. We focus
  on the gauge-invariant contribution mediated by a loop of quarks. We
  explicitly include all effects of the loop-quark mass $m$,
  appropriate for the production of a pair of $Z$-bosons. In the limit
  $m \to 0$, our results are also applicable to the production of
  $W$-boson pairs, mediated by a loop of massless quarks. Implemented
  in a numerical code, the results are fast. The calculation
  uses novel advancements in spinor-helicity simplification
  techniques, for the first time applied beyond five-point massless
  kinematics. We make use of primary decompositions from
  algebraic-geometry, which now involve non-radical ideals, and
  $p$-adic numbers from number theory. We show how to infer whether
  numerator polynomials belong to symbolic powers of non-radical ideals
  through numerical evaluations.}


\begin{document} 
\setcounter{tocdepth}{2}
\maketitle

\section{Introduction}

In many respects the numerical calculation of one-loop amplitudes, both in the
standard model and in proposed models beyond the standard model, is a solved problem. Following Passarino and
Veltman~\cite{Passarino:1978jh} the problem is separated into the
calculation of scalar one-loop integrals and the calculation of the coefficients with
which these integrals appear in the particular amplitude at hand.  The
needed finite scalar integrals are provided in
ref.~\cite{tHooft:1978jhc,Denner:1991qq}, whereas the needed singular integrals are
provided in ref.~\cite{Ellis:2007qk}.  Techniques based on numerical
unitarity~\cite{Ossola:2006us,Ellis:2011cr} as well as methods based on iterative
calculation of Feynman diagrams~\cite{Cascioli:2011va,Buccioni:2019sur,Denner:2017wsf}
have been automated into sophisticated tools, which give reliable
numerical results for the coefficients of the scalar integrals.

On the other hand, analytic unitarity techniques have also matured so
they can give analytic results following an automatic
recipe~\cite{Britto:2004nc,Forde:2007mi,Badger:2008cm,Mastrolia:2009dr,Ellis:2011cr}.
However, the resultant analytical expressions are often complicated so
that the singularity structure of the amplitude is hard to divine and
numerical evaluations are suboptimal. In this paper we push the
analytic techniques a step further by obtaining simpler analytic
expressions where the form of the answer, especially with regard to
the singularity structure, is manifest. The possible benefits of
utilizing the amplitude in such a form are:
\begin{itemize}
\item the simpler form may lead to faster numerical evaluation;
\item the singularity structure is manifest, with physical poles of as low
degree as possible;
\item the consequent analytic form leads to improved stablility of
  numerical evaluation;
\item the numerical behaviour around singular points can be improved
  by analytic expansions, if necessary.
\end{itemize}

The issue of stable evaluation of the amplitude is particular pressing
for the case of vector boson pair production. Cuts on the
transverse momenta of the decay products of the
vector bosons do not exclude the region where the vector sum of the
transverse momenta of all the uncoloured particles in the final state is equal to
zero. In this kinematic region the amplitudes contain soft and collinear singularities.
This issue is especially important in the context of next-to-next-to-leading
order calculations since cancellations between real and virtual diagrams
occur in the region of zero transverse momentum.

A compact representation of scattering amplitudes is provided in principle
by the spinor-helicity formalism~\cite{Berends:1981uq,Xu:1986xb,Dixon:2013uaa}.
These are especially convenient for the case at hand because
the factors associated with vector boson decays are simple and the whole family
of diboson processes may be described by appropriate dressings of a core set of
amplitudes~\cite{Dixon:1998py}. However,
since the spinor products are not all independent, spinor
product expressions are not straightforward to simplify.  In particular, the application
of momentum conservation and Schouten identities leads to many equivalent representations
of the same amplitude. The application of systematic reduction techniques (such as momentum twistors
or Gr\"obner basis reduction) does not necessarily result in simpler expressions.

A number of strategies are available to facilitate simplification.
The method of momentum twistors~\cite{Hodges:2009hk,Badger:2013gxa} allows one to write spinor expressions
contributing to an $n$-point amplitude in a unique form in terms of $3n-10$ independent variables. However it is
cumbersome to revert to simple spinor expressions as the number of external legs grows.
Alternatively, simpler expressions may be obtained by reconstructing
multivariate polynomials and rational functions from their evaluation over finite fields~\cite{vonManteuffel:2014ixa,Peraro:2016wsq,Maierhofer:2017gsa,Smirnov:2019qkx,Klappert:2019emp}.

Our analytic results will be simplified using large-precision floating point arithmetic and fitting in singular limits~\cite{DeLaurentis:2019phz}, as well as using $p$-adic numbers and technology from algebraic-geometry~\cite{DeLaurentis:2022otd}.
The expressions we provide are explicitly rational, i.e.~no square roots are present, and
contain poles of the lowest degree possible.

\subsection{Motivation}
Our motivation for this paper is twofold. First, we want to
investigate and extend the practical limits on the numerical
simplification techniques, alluded to above and to be explained in
more detail in Section~\ref{Geometric}. In particular, it is
interesting to investigate their feasibility in high multiplicity
settings, where new poles appear in the master-integral
coefficients. For this purpose, we choose a subset of the diagrams
contributing to the one-loop 7-point process $q \bar{q} \ell \bar\ell
\ell^\prime \bar{\ell}^\prime g$, namely the diagrams including a
quark loop.  Second, vector boson pair production is an important
process, to which these amplitudes contribute. For example, the
$ZZ$ final state is one of the decay channels of the Higgs boson, and
onshell and offshell calculations are of great interest. The
amplitudes we consider contribute to,
\beq
\label{procZZ}
\bothdk{q + \bar{q}}{Z/\gamma^*+}{Z/\gamma^*+g}{\ell^\prime \, \bar\ell^\prime}{\ell \, \bar\ell}
\eeq
and to the related processes obtained by crossing the coloured partons.
This amplitude receives contributions both at tree level and at one loop.
In this paper we report on the one-loop amplitude mediated by quarks with a common mass, $m$.
In the limit $m \to 0$ our results can also be used for the massless quark-loop contributions to
the process,
\beq
\bothdk{q + \bar{q}}{W^-+}{W^++g}{\nu^\prime \bar{\ell}^\prime}{\ell\, \bar{\nu}} 
\eeq
We note that results for this process, in the limit of massless quarks circulating in the loop,
have been calculated in analytic form previously~\cite{Campbell:2015hya}, but
the resulting amplitudes were only distributed in the form of computer
code since they were not sufficiently compact to present otherwise.

In this paper we will present simpler formulae for all the ingredients necessary to assemble
the quark-loop amplitude for the process in Eq.~\eqref{procZZ}.
In addition, we have added the simplified results to the code
MCFM~\cite{Campbell:1999ah,Campbell:2011bn,Boughezal:2016wmq}, which we will demonstrate
leads to improvements in speed.

\subsection{Plan of this paper}
In Section~\ref{lowestorder} we review the spinor notation and use it to present results for the lowest order
amplitude. 
In Section~\ref{Geometric} we review and build upon the algebro-geometric methods of ref.~\cite{DeLaurentis:2022otd}, with additional details worked out in appendix \ref{sec:AsymmetricApproachesAndRingExtensions}.
Section~\ref{Integral_Coefficients} (Appendix~\ref{anomtriangles}) presents the analytic results for the coefficients of box, triangle and bubble scalar integrals
coming from quark-loop box (triangle diagrams) contributing to the process in Eq.~\eqref{procZZ}.
Timing results from the numerical implementation of our calculation are described in Section~\ref{timing} and
our conclusions are given in Section~\ref{conclusions}.

\section{Lowest order amplitude}
\label{lowestorder}
\subsection{Spinor notation}
\label{spinorsection}
We begin this section by introducing the notation which we shall use to present the tree-graph results
for the $Z$-pair production amplitude, as well as for the one-loop results to be
presented in Section~\ref{Integral_Coefficients}.  
All results are presented using the standard notation for the kinematic invariants of the process,
\beq
s_{ab} = (p_a+p_b)^2 \, ,
s_{abc} = (p_a+p_b+p_c)^2 \, ,
s_{abcd} = (p_a+p_b+p_c+p_d)^2 \,,
\end{equation}
and the Gram determinant,
\beq \label{Delta3eqn}
\DeltaThree(a,b,c,d) =(s_{abcd}-s_{ab}-s_{cd})^2-4 s_{ab} s_{cd}  \, .
\eeq
The Weyl spinor $\lambda_A$ is a two dimensional complex vector and its complex conjugate
is denoted as $\bar{\lambda}_{\dot{A}}$. The spinorial inner product between to Weyl spinors 
$\lambda_1$ and $\lambda_2$ is written as,
\beqn
\lambda_{1\,A}\lambda_{2\,B} \varepsilon^{BA} &=& \lambda_{1\,A}\lambda_{2}^{A} = \langle 12 \rangle \, ,\\
\bar{\lambda}_{1\,\dot{A}} \bar{\lambda}_{2\,\dot{B}}\varepsilon^{\dot{B}\dot{A}} &=& \bar{\lambda}_{1\,\dot{A}}\bar{\lambda}_{2}^{\dot{A}} = [12]  \, ,
\eeqn
where $\varepsilon$ is the totally antisymmetric tensor in two dimensions,
\beq
\varepsilon_{AB}=\varepsilon^{AB}=\varepsilon_{\dot{A}\dot{B}}=\varepsilon^{\dot{A}\dot{B}}=
\left(\begin{matrix} 0 & +1 \\ -1 & 0\end{matrix}\right)\, .
\eeq
For a light-like momentum $p$, (i.e. $p^2=(p^0)^2-(p^1)^2-(p^2)^2-(p^3)^2=0$)
we have that,
\beq
\lambda_{a\,A} \equiv \langle a|=\left( \begin{matrix}\sqrt{p_a^+}\exp(+i\varphi_{p_a}), & \sqrt{p_a^-}\end{matrix}\right),\;\;
\lambda_b^A \equiv | b \rangle=\left( \begin{matrix}-\sqrt{p_b^-}\\ \sqrt{p_b^+}\exp(+i\varphi_{p_b})\end{matrix}\right),
\eeq
where 
\beq \label{eq:phasekdef}
e^{\pm i\varphi_p}\ \equiv\ 
  \frac{ p^1 \pm ip^2 }{ \sqrt{(p^1)^2+(p^2)^2} }
\ =\  \frac{ p^1 \pm ip^2 }{ \sqrt{p^+p^-} }\ , \qquad p^\pm\ =\ p^0 \pm p^3.  
\eeq
For the conjugate spinor we have that,
\beq
\bar{\lambda}_{a \,\dot{A}}=[a|=\left( \begin{matrix}\sqrt{p_a^+}\exp(-i\varphi_{p_a}),& \sqrt{p_a^-}\end{matrix}\right),\;\;
\bar{\lambda}_b^{\dot{A}}=|b]=\left( \begin{matrix}-\sqrt{p_b^-}\\ \sqrt{p_b^+}\exp(-i\varphi_{p_b})\end{matrix}\right)\, .
\eeq
The corresponding results for the spinor products are
\beqn
\label{Spinor_products1}
\lambda_{a\,A} \lambda_b^A&=&\langle ab \rangle=\sqrt{p_a^- p_b^+}\exp(+i\varphi_{p_a})-\sqrt{p_a^+ p_b^-}\exp(+i\varphi_{p_b})\, , \\
\label{Spinor_products2}
\bar{\lambda}_{a\,\dot{A}} \bar{\lambda}_b^{\dot{A}}&=&[ab]=\sqrt{p_a^+ p_b^-}\exp(-i\varphi_{p_b})-\sqrt{p_a^- p_b^+}\exp(-i\varphi_{p_a})\, .
\eeqn
The spinor products are antisymmetric, $\langle ba \rangle = -\langle ab \rangle$, $[ba] = -[ab]$
and since the four-vectors are light-like, $s_{ab}=\langle ab\rangle [ba]$.
The spinor products also satisfy the Schouten identity,
\beq
\label{eq:SchoutenIdentity}
\spa a.b \spa c.d = \spa a.d \spa c.b + \spa a.c \spa b.d,\;\;\;
\spb a.b \spb c.d = \spb a.d \spb c.b + \spb a.c \spb b.d\, .
\eeq

In order to write the amplitude in
a simple form we introduce some extra notation for more complicated spinor sandwiches.
We have for light-like vectors $a \ldots h$,
\begin{eqnarray}
\label{Spinor_products_long}
\spab{a}.{bc}.{d}&=& \spa{a}.{b} \spb{b}.{d} +\spa{a}.{c} \spb{c}.{d} \, ,\nonumber \\
\spaba{a}.{bc}.{de}.{f} &=& \spab{a}.{bc}.{d} \spa{d}.{f} +\spab{a}.{bc}.{e}\spa{e}.{f} \, ,\nonumber \\
\spabab{a}.{bc}.{de}.{fg}.{h} &=& \spaba{a}.{bc}.{de}.{f} \spb{f}.{h}+\spaba{a}.{bc}.{de}.{g} \spb{g}.{h}\, ,\nonumber\\
\spba{a}.{bc}.{d} &=& \spb{a}.{b} \spa{b}.{d} +\spb{a}.{c} \spa{c}.{d}\, ,  \nonumber \\
\spbab{a}.{bc}.{de}.{f} &=& \spba{a}.{bc}.{d} \spb{d}.{f}+\spba{a}.{bc}.{e}\spb{e}.{f}\, ,\nonumber \\
\spbaba{a}.{bc}.{de}.{fg}.{h}&=& \spbab{a}.{bc}.{de}.{f}\spa{f}.{h}+\spbab{a}.{bc}.{de}.{g}\spa{g}.{h}\, .
\end{eqnarray}
We also define $\tilde\Gamma$ (note the tilde) to imply anti-symmetrization,
\begin{eqnarray}\label{eq:gamma_tilde}
  \langle a| \tilde{\Gamma}_{bc|de}  |f \rangle &=& \spaba{a}.{bc}.{de}.{f} - \spaba{a}.{de}.{bc}.{f} \, , \nonumber \\
  \lbrack a| \tilde{\Gamma}_{bc|de}  |f \rbrack &=& \spbab{a}.{bc}.{de}.{f} - \spbab{a}.{de}.{bc}.{f} \, , \nonumber \\
  \lbrack a| \tilde{\Gamma}_{bc|de|fg} |h\rangle & = & \spbaba{a}.{bc}.{de}.{fg}.{h} - \spbaba{a}.{fg}.{de}.{bc}.{h} \, .
\end{eqnarray}
Lastly, we point out that Eq.~\eqref{Spinor_products_long} and Eq.~\eqref{eq:gamma_tilde}
imply definitions for non-fully contracted spinor strings. These un-contracted
spinor strings are useful to express generators with open indices for some ideals in Section~\ref{codim2}.

The basic amplitude which we calculate is the one for the process involving four leptons, two quarks and one gluon, 
\beq \label{momentumlabelling}
A_7(1_q^-,\,2_{\qbar}^+,\,3_\ell^-,\,4_{\lbar},\,5_{\ell^\prime}^-,\,6_{\lbar^\prime}^+,\,7_g^+) \, ,
\eeq
with all particles outgoing; the superscript denotes the helicity and the subscript shows the type of particle.
The couplings required to reconstruct the physical amplitude will be given later.

\subsection{Tree graphs}
We will first setup the notation for the reduced amplitudes
removing the colour matrix (for emission of a gluon with colour index $B$),
and powers of the coupling constants,
\beq \label{colourstrip}
    {\cal A}_7^{\tree,B}(1,2,3,4,5,6,7)= 4 i  g_s e^4 (t^B)_{{i_1 \ib_2}} A_7^{\tree}(1,2,3,4,5,6,7)\, .
\eeq
The colour matrix $t^B$ is normalized such that ${\rm tr}(t^A t^B) = \delta^{AB}$. The indices
$i_1$ and $\ib_2$ denote the colours of the quark and anti-quark line, $i_1,\ib_2=\{1,2,3\}$.
The strong and electromagnetic couplings are denoted by $g_s$ and
$e$. Written in this form ${\cal A}_7^{\tree,B}$ is exactly the amplitude
where the production of both pairs of leptons (off a unit electric
charge quark line) is mediated by virtual photons.  The appropriate
coupling factors and propagators for $Z$ boson production will be
added below.  The result for the reduced tree amplitude is,

\beqn
\label{bremsampA}
A_7^{\tree}(1^-,2^+,3^-,4^+,5^-,6^+,7^+) &=&
-\frac{\spa1.3}{\spa1.7 s_{34}\,s_{56}\,s_{134}} \\
&\times&\left[
   \frac{\spa1.3\spb3.4\spb2.6\spab5.26.7}{s_{256}}
 + \frac{\spab5.13.4 \spab1.27.6 }{\spa7.2} \right]\nonumber \, .
\eeqn
This amplitude was first presented in ref.~\cite{Dixon:1998py} although our labelling of the
lepton momenta, see Eq.~\eqref{momentumlabelling}, differs from the notation in that paper.
The remaining tree amplitudes are obtained by symmetry operations.  These correspond to
flipping the helicities of the leptons, e.g.
\beq 
A_7^{\tree}(1^-,2^+,3^+,4^-,5^-,6^+,7^+) =  A_7^{\tree}(1^-,2^+,4^-,3^+,5^-,6^+,7^+) \,,
\eeq
flipping the helicities of the quarks, e.g.,
\beq 
A_7^{\tree}(1^+,2^-,3^-,4^+,5^-,6^+,7^+) = A_7^{\tree}(2^-,1^+,3^-,4^+,5^-,6^+,7^+) \,,
\label{eq:drquarkflip}
\eeq
and reversing the helicity of the gluon, e.g.,
\beq 
A_7^{\tree}(1^+,2^-,3^-,4^+,5^-,6^+,7^-) = \left. - A_{sr}(1^-,2^+,4^-,3^+,6^-,5^+,7^+)
 \right|_{\spa.. \leftrightarrow \spb..} \,.
\eeq

\subsubsection{Restoring the couplings}
We define the left- and right-handed couplings of a $Z$ boson to quarks and
leptons by,
\begin{eqnarray}
\label{quarkcouplings}
&&v_{L,q} = \frac{\tau_q - 2 Q_q \sstw}{2\stw\ctw}, \qquad
v_{R,q} = - \frac{Q_q \stw}{\ctw}, \\
\label{leptoncouplings}
&&v_{L,e} = \frac{-1 + 2 \sstw}{2\stw\ctw}, \qquad \;\;\,
  v_{R,e} = \frac{\stw}{\ctw}, \\
\label{leptoncouplings2}
&&v_{L,n} = \frac{1}{2\stw\ctw}, \qquad \;\;\,
  v_{R,n} = 0,
\end{eqnarray}
where $Q_q$ is the charge of the quark (in units of the positron charge) and
$\tau_q = +1$ for up-type quarks and $\tau_q = -1$ for down-type quarks.
The full tree amplitude for the $ZZ$ case is then,
\begin{eqnarray}
\label{tree_amplitude_with_couplings}
{\cal A}_7^{\tree,B}(1,2,3,4,5,6,7)
 &=& 4 i e^4 g_s (t^B)_{{i_1 \ib_2}} \nn \\
 &\times& \left( q_{34} Q_q + v_{34} v_q P(s_{34},M_Z) \right)
  \left( q_{56} Q_q + v_{56} v_q P(s_{56},M_Z) \right) \nn \\
 &\times& \left( A_7^{\tree}(1,2,3,4,5,6,7) + A_7^{\tree}(1,2,5,6,3,4,7) \right) \, ,
\end{eqnarray}
where $q_{34}$, $v_{34}$ ($q_{56}$, $v_{56}$) label the charge and coupling
factors for the leptons appearing in the decay of $Z(p_{34})$ ($Z(p_{56})$)
as given in Eqs.~\eqref{leptoncouplings} and~\eqref{leptoncouplings2}.
The $Z$-boson propagator factor is given by,
\beq
P(s,M)=\frac{s}{s-M^2}\, ,
\eeq
where $M$ is the (complex) mass of the vector boson.
Dressed this way, these amplitudes account for the effect of virtual photons
as well as $Z$ bosons.

For completeness, we also present the tree graph results for the contribution of singly-resonant diagrams
$Z(p_{3456}) \to \ell_3 \bar\ell_4 Z(p_{56})$.  The colour stripped amplitudes, reduced as in Eq.~\eqref{colourstrip},
for this contribution are,
\begin{eqnarray}
&& {A}_7^{sr}(1^-,2^+,3^-,4^+,5^-,6^+,7^-) = \nn \\ &&
\left(\frac{\spa3.5 \spb2.4 \spbab2.17.35.6}{s_{356}}    
   +\frac{\spb6.4 \spab3.17.2 \spab5.46.2}{s_{456}}\right)
  \frac{1}{\spb2.7 \spb7.1 s_{3456} s_{56}} \, ,
\end{eqnarray}
and,
\begin{eqnarray}
&& {A}_7^{sr}(1^-,2^+,3^-,4^+,5^-,6^+,7^+) = \nn \\ &&
  \left(\frac{\spa1.3 \spb6.4 \spaba5.46.27.1}{s_{456}}
  +\frac{\spa5.3 \spab1.35.6 \spab1.27.4}{s_{356}}\right)
  \frac{1}{\spa2.7 \spa1.7 s_{3456} s_{56}}\, .
\end{eqnarray}
The quark helicities can be flipped by, for example,
\beq 
A_7^{sr}(1^+,2^-,3^-,4^+,5^-,6^+,7^+) = -A_7^{sr}(2^-,1^+,3^-,4^+,5^-,6^+,7^+) \,,
\eeq
where we note that there is an additional sign-flip compared to the corresponding relation
for the double-resonant contribution in Eq.~\eqref{eq:drquarkflip}.
Amplitudes with lepton helicities $(5^+, 6^-)$ are obtained by the interchange $5 \leftrightarrow 6$.
For lepton helicities $(3^+, 4^-)$ we have, for example,
\beq 
A_7^{sr}(1^+,2^-,3^+,4^-,5^+,6^-,7^+) = \left. - A_7^{sr}(1^-,2^+,3^-,4^+,5^-,6^+,7^-)
 \right|_{\spa.. \leftrightarrow \spb..} \,.
\eeq
The contribution of the singly resonant diagrams to the full amplitude,
dressed with couplings and adding also the term with the
role of the $Z$ bosons interchanged, is given by,
\begin{eqnarray}
&&{\cal A}_7^{\tree,sr,B}(1,2,3,4,5,6,7) = 4 i e^4 g_s (t^B)_{{i_1 \ib_2}} \Big[ \\ &&
   \left( q_{34} q_{56} + v_{34} v_{56} P(s_{56},M_Z) \right)
   \left( q_{34} Q_q + v_{34} v_q P(s_{3456},M_Z) \right)
 \, A_7^{sr}(1,2,3,4,5,6,7) \nn \\ 
&+&\left( q_{34} q_{56} + v_{34} v_{56} P(s_{34},M_Z) \right)
   \left( q_{56} Q_q + v_{56} v_q P(s_{3456},M_Z) \right)
 \, A_7^{sr}(1,2,5,6,3,4,7) \Big] \nn \, .
\end{eqnarray}

\section{Geometric Reconstruction}
\label{Geometric}

Any one-loop amplitude can be written as a sum of scalar master
integrals with definite coefficients and rational
terms~\cite{Passarino:1978jh}. Therefore, we write,
\begin{eqnarray} \label{scalarreduction0}
 A_7^{\oneloop} &\sim&
 \sum_{i,j,k} {d}_{\{i\sep j\sep k\}} \, D_0(p_i, p_j, p_k ;m)  
+ \sum_{i,j} {c}_{\{i\sep j\}} \,  C_0(p_i,p_j ;m)   \nonumber \\
 &+& \sum_{i} {b}_{\{i\}} \, B_0(p_i;m)  + {r}\, .
\end{eqnarray}
The exact definitions of the scalar integrals $D_0$, $C_0$, and $B_0$
are given in Appendix~\ref{Integrals}. For the case at hand, the
rational term is fully determined by the mass dependence of the
coefficients~\cite{Badger:2008cm}. In this section, we deal with the
problem of simplifying the coefficients of the master integrals:
${d}_{\{i\sep j\sep k\}}$, ${c}_{\{i\sep j\}}$ and ${b}_{\{i\}}$. Even
though the coefficients are obtained through analytic unitarity
methods, we phrase the simplification problem as a reconstruction
problem from numerical samples.  In this context, we treat all three
types of coefficients on the same footing, thus we give them a generic
name $\mathcal{C}_i$. These coefficients $\mathcal{C}_i$ are rational
functions of the external kinematics. We write
\beq
  \mathcal{C}_i(\lambda, \tilde\lambda) = \frac{\mathcal{N}_i(\lambda, \tilde\lambda)}{\prod_j \mathcal{D}_j(\lambda, \tilde\lambda)^{q_{ij}}} \, ,
\eeq  
where $(\lambda, \tilde\lambda)$ denote the set of right- and
left-handed Weyl spinors, which we treat as independent. That is, we
consider the $\mathcal{C}_i$ in the analytical continuation to complex
momenta. The exponents $q_{ij}$ are integers and they are allowed to
take negative values, thus denoting common factors in the
numerator. From a geometric perspective, since these rational
coefficients are functions of many complex variables, they are evaluated 
in a multi-dimensional space analogous to the complex plane. Their poles and
zeros will be surfaces (a.k.a.~\textit{varieties}) of one less
dimension than the full space. Leveraging the geometric picture, in
this section we describe the approach employed to simplify the
amplitudes presented in this paper.

In our theoretical approach we make use of elements of both the
algebro-geometric reconstruction procedure of
ref.~\cite{DeLaurentis:2022otd} and of the iterated, in-limit
reconstruction strategy of ref.~\cite{DeLaurentis:2019phz}. On a more
practical level, we rely on the computer algebra system
\texttt{Singular} \cite{DGPS}, through the \texttt{Python} interface
\texttt{syngular} \cite{syngular}, on the multi-precision
floating-point arithmetic package \texttt{mpmath} \cite{mpmath} and on
an in-house \texttt{Python} implementation of $p\verytiny$-adic
numbers $(\mathbb{Q}_p)$ with variable size mantissa and explicit
precision tracking.

We begin in Section~\ref{background} with a brief review of the most
important algebro-geometric concepts, namely ideals and varieties in
spinor space. The aim here is mainly to set up notation and recall
concepts; for a more comprehensive discussion we refer the reader to
ref.~\cite{DeLaurentis:2022otd}, and references therein. In Section~\ref{codim1},
we present the poles and zeros of the coefficients as
varieties in spinor space, with an associated degree of divergence or
vanishing. Then, in Section~\ref{codim2} we consider intersections of
these varieties, and their decompositions into irreducible
components. Numerical evaluations close to these new,
lower-dimensional varieties are used to obtain constraints on the
numerator structure and, in particular, on the possible partial
fraction decompositions. Finally, we sample the coefficients near
singular varieties to rationally reconstruct the remaining free
parameters of the ansatz.

\subsection{Ideals and varieties in spinor space}
\label{background}

For the purposes of algebro-geometric computations, let us begin by
considering polynomials in the components of the Weyl spinors of the
massless external legs.  Mathematically, we say that these polynomials
belong to the polynomial ring defined as,
\beq\label{eq:polynomial_ring}
S_n = \mathbb{F}\big[ |1⟩, [1|, \dots, |n⟩, [n| \big] \, ,
\eeq
where the spinors are understood to be taken component-wise and we
employ standard spinor-helicity notation. We stress that, in the
analytic continuation to complex momenta, the spinors are 
independent. The field $\mathbb{F}$ is taken to be either the complex
numbers ($\mathbb{C}$) or the $p\verytiny$-adic numbers
($\mathbb{Q}_p$). In practice, we usually work over $\mathbb{C}$ and
all geometric considerations are always understood to be over
the complex numbers, as it is often required for the field to be
algebraically closed. However, in some circumstances evaluations
over $\mathbb{Q}_p$ can be very useful since the scale hierarchy
is easier to control.

In order to discuss the geometric properties of the amplitude
coefficients, we introduce the algebraic concept of an ideal, which is
denoted by a pair of angle brackets\footnote{Unfortunately, this
common notation from algebraic geometry clashes with standard
spinor-helicity notation. Nevertheless, all expressions are
unambiguous: the brackets denoting ideals must be balanced.}.  An
ideal is defined to be the set of all polynomial linear combinations
of an initial set of polynomials which are called generators.  As
algebraic objects, ideals have well defined algebraic operations,
addition, multiplication etc.  In particular, a physically important
ideal of $S_n$ is that of momentum conservation,
\beq
\label{eq:mom_cons}
  J_{\Lambda_n} = \Big\langle \sum_{i=1}^n |i⟩[i| \Big\rangle_{S_n} \, ,
\eeq      
where the subscript denotes the ring to which the ideal belongs.  Note
that, we can write this ideal as a single tensor generator,
or as four generators taking the tensor component by component.
If instead, we were to define the momentum conservation ideal in the subring of spinor products
(see Eqs.~\eqref{Spinor_products1} and \eqref{Spinor_products2} and
ref.~\cite[Section 2.2]{DeLaurentis:2022otd})
it would require $n^2$ contractions of momentum conservation from
Eq.~\eqref{eq:mom_cons}, plus an additional $2 {n \choose 4}$ Schouten
identities from Eq.~\eqref{eq:SchoutenIdentity}.

For the spinors to describe a physically meaningful phase space, they
must statisfy momentum conservation. That is, any polynomial belonging
to the momentum conservation ideal $J_{\Lambda_n}$ has to be
considered a rewriting of zero. For this reason, it is convenient to
introduce the quotient ring,
\beq
\label{eq:quotient_ring}
  R_n = S_n / J_{\Lambda_n} \, .
\eeq
This is the set of all polynomials in spinor components where any pair
of polynomials differing by a member of $J_{\Lambda_n}$ is considered
to be equivalent. That is, the elements of $R_n$ are not polynomials, but
equivalence classes of polynomials. For example, the following are six different ways
to write the same element of $R_7$,
\beq
\Big\{\, \spbab7.12.34.7, \spbab7.34.56.7, \spbab7.56.12.7, -\spbab7.34.12.7, -\spbab7.56.34.7, -\spbab7.12.56.7 \, \Big\} \, .
\eeq
Just like we can define ideals of $S_n$, we can also define ideals of
$R_n$. In fact, there exists a one-to-one map between ideals of $R_n$
and ideals of $S_n$ that contain $J_{\Lambda_n}$, given by,
\beq\label{eq:qring-ideal-correspondence}
\Big\langle \, p_1, \dots, p_k \, \Big\rangle_{R_n} \sim \Big\langle \, p_1, \dots, p_k, \sum_{i=1}^n |i⟩[i| \, \Big\rangle_{S_n}
\eeq
for an arbitrary set of generators $p_1,\, \dots, \, p_k$.  The
numerators $\mathcal{N}_i$ and the denominators $\prod_j
\mathcal{D}_j^{q_{ij}}$ of the coefficients $\mathcal{C}_i$ belong to
$R_n$. Therefore, the coefficients $\mathcal{C}_i$ belong to the field
of fractions of $R_n$, denoted as $FF(R_n)$ . However, note that it is
not entirely trivial to define a field of fractions over a quotient
ring (see ref.~\cite[Chapter 5]{cox2006using}). In particular, $R_n$
needs to be a so-called integral domain, and $R_3$ is
not\footnote{This follows from the fact that $\big\langle 0
  \big\rangle_{R_3}$ is not a prime ideal (see below).}. For $n\geq
4$, $R_n$ is an integral domain and $FF(R_n)$ is well defined. We will
work in $R_7$.
    
The key geometric concept that we consider is that of a variety. We
denote a variety $U$ associated to an ideal $J$ as $U=V(J)$.  For an
ideal $J$ of $S_n$, $V(J)$ is defined as the set of points
$(\lambda, \tilde\lambda) \in \mathbb{F}^{4n}$ such that the generators of $J$
evaluate to zero. Through the correspondence of
Eq.~\eqref{eq:qring-ideal-correspondence}, the same definition applies
to varieties associated to ideals of $R_n$. Note that in the latter
case, all varieties will be sub-varieties of
$V(J_{\Lambda_n})$. Similarly, to every variety $U$ we can associate
an ideal $J$, denoted as $J=I(U)$, and defined as the set of
polynomials which vanish on $U$. Varieties and ideals have well
defined dimensions. For example, $\text{dim}(V(J_{\Lambda_n}))=(4n-4)$,
as 4 constraints are imposed in $\mathbb{F}^{4n}$. We also define
codimension as the complement of dimension w.r.t.~the dimension of the
full space, i.e.~$\text{codim}(V(J_{\Lambda_n}))=4$. Note that the
codimension need not always match the number of generators. If it is
possible to find a set of generators with as few elements as the
codimension, then we describe the ideal as of maximal codimension. As
the quotient of a polynomial ring by a maximal codimension ideal,
$R_n$ is a Cohen--Macaulay ring \cite{DeLaurentis:2022otd}. This
property of $R_n$ has a number of useful implications which we will
review shortly. In the context of $R_n$, as $V(\big\langle 0
\big\rangle_{R_n}) \sim V(J_{\Lambda_n})$, a codimension-one variety
will be a $[(4n-4)-1]$-dimensional variety contained in
$V(J_{\Lambda_n})$ and a codimension-two variety will have dimension
$[(4n-4)-2]$. The poles of the rational coefficients are codimension
one varieties.

It turns out that the order of a pole is not necessarily well defined
on all varieties.  If a variety is comprised of multiple irreducible
components, which we call branches, then the degree of divergence need
not be the same for all of them. Therefore, it is important to be able
to identify the branches of reducible varieties. Any variety admits a
unique minimal decomposition,
\beq
  \label{eq:minimal-decomposition-variety}
  U = \bigcup_{k=1}^{n_B(U)} U_k \, ,
\eeq
where $n_B(U)$ denotes the number of branches $U_k$. If $U$ itself is
irreducible then we simply have $n_B(U)=1$.  To compute a minimal
decompositions of a variety, one can rely on the corresponding concept
for ideals, that is on a so-called minimal primary
decomposition. Given any ideal $J$, in analogy to
Eq.~\eqref{eq:minimal-decomposition-variety}, we can write,
\beq
  \label{eq:minimal-primary-decomposition}
  J = \bigcap_{l=1}^{n_Q(J)}Q_l \, ,
\eeq  
where $n_Q(J)$ denotes the number of primary ideals $Q_l$ in the
primary decomposition. Each primary ideal $Q_l$ has an associated
prime ideal $P_l=\sqrt{Q_l}$, where the root denotes the ideal
radical. One says that $Q_l$ is $P_l$-primary. Now let $U = V(J)$. The
subset of prime ideals $P_l$ such that $P_l=V(U_k)$ for some $k$ is
called the set of minimal associated primes. It has $n_B(U)$ elements
and we denote it as $\text{minAssoc}(J)$. The complement of this subset
is the set of so-called embedded components. An embedded prime $P_l$
is such that $V(P_l)$ is redundant (and thus absent) in
Eq.~\eqref{eq:minimal-decomposition-variety}, but such that $Q_l$ is
not redundant in Eq.~\eqref{eq:minimal-primary-decomposition}. The
number of embedded components is $n_Q(J)-n_B(U)$. A first useful
consequence of $R_n$ being a Cohen--Macaulay ring is that maximal
codimension ideals in $R_n$ are equi-dimensional, i.e.~they are free
of embedded components \cite{DeLaurentis:2022otd},
\beq\label{eq:first-CM-consequence}
J = \big \langle p_1, \dots, p_m \big \rangle_{R_n} \quad \text{s.t.}
 \quad \text{codim}(J) = m \quad \Longrightarrow \quad n_Q(J) = n_B(V(J)) \, .
\eeq
It is also useful to note that $S_n$, $J_n$ and $R_n$ are symmetric
under permutations of the external legs and a swap of left- and
right-handed spinors. We can make use of these properties to aid the
computation of primary decompositions.

Since the degree of divergence of a rational fraction of $R_n$ when
considered simultaneously near a pair of poles is often less than the
sum of the divergences near each pole separately, it is useful to
consider elements of $R_n$ vanishing to a given order on a given
variety. Given an irreducible variety $V(Q_l)$ and a numerator
$\mathcal{N}_i$ which vanishes to order $\kappa$ on $V(Q_l)$, the
Zariski-Nagata theorem \cite{Zariski:1949,Nagata1962,EISENBUD1979157}
tells us that $\mathcal{N}_i$ has to belong to the
$\kappa^{\text{th}}$ symbolic power of the associated prime ideal
$P_l$, denoted as $P_l^{\langle \kappa \rangle}$. This power $\kappa$
can reliably be identified numerically \cite{DeLaurentis:2022otd}. To
see why a refined notion of power is needed, let us consider the
standard ideal power $P_l^\kappa$ of a prime ideal $P_l$. It is
defined through repeated ideal multiplication and it may not be a
primary ideal, i.e.~$P_l^\kappa$ may involve embedded
components. These embedded components imply additional non-trivial
vanishing properties on sub-varieties of $V(Q_l)$. In contrast, the
symbolic power $P_l^{\langle \kappa \rangle}$ is defined as the
$P_l$-primary component of $P_l^\kappa$. A second useful consequence
of $R_n$ being a Cohen--Macaulay ring is that symbolic powers of
maximal codimension ideals in $R_n$ coincide with the standard powers,
\beq\label{eq:second-CM-consequence}
J = \big \langle p_1, \dots, p_m \big \rangle_{R_n} \quad \text{s.t.} \quad \text{codim}(J) = m \quad \Longrightarrow \quad J^{\langle \kappa \rangle} = J^\kappa \, .
\eeq

\subsection{Poles and zeros as codimension-one varieties}\label{codim1}

We begin by determining the denominators of the coefficients
$\mathcal{C}_i$, which belong to $FF(R_7)$. Poles and zeros of these
rational functions are irreducible varieties of codimension one, with
an associated degree of vanishing or divergence. we observe that the poles needed for
the one-loop amplitudes under our consideration in this paper are of the form,
\beq\label{eq:poles-list}
\mathcal{D} = \big\{ \langle ab\rangle, \, \langle a|\Gamma_{bc}|a], \,
        \langle a|\Gamma_{bc|de}|a\rangle, \, \Delta_3(a,b,c,d) \big\} \, ,
\eeq
where the indices $a, b, c, d, e$ are assumed to be distinct and in
the set $\{1, \dots, 7\}$.  All associated codimension-one ideals,
that is ideals of the form $\langle D_j \rangle_{R_7}$, are prime,
meaning that the corresponding varieties $V(\langle D_j
\rangle_{R_7})$ are irreducible. Then, it follows that the order of
the poles and zeros can be determined by either a pair of high-precision
floating point evaluations in the limit approaching the variety, as by
ref.~\cite{DeLaurentis:2019phz}, or by a single $p\verytiny$-adic
evaluation at a point close to the variety, as by
ref.~\cite{DeLaurentis:2022otd}. With this procedure we obtain the least
common denominators (LCD),
\beq
\mathcal{D}_{\text{LCD},\,i} = \prod_j \mathcal{D}_j(\lambda, \tilde\lambda)^{q_{ij}} \; , \; \text{with} \;\; q_{ij} > 0 \,.
\eeq
We stress that since the pole orders are determined numerically, it
follows that all spurious poles are automatically removed and the
physical ones are of as low degree as posisble. Common factors in the
numerator ($q_{ij}<0$) are also obtained in this way.

\subsection{Partial fractions and numerators from codimension-two varieties}\label{codim2}

To proceed in the simplification we aim to constrain the numerators
$\mathcal{N}_i$. For this purpose we study of their behaviour on
codimension-two varieties. In particular, it is convenient to consider
the behaviour of the numerators on those codimension-two varieties that
originate from the intersection of varieties of codimension one
corresponding to poles of the coefficients. We can now assume we have
access to numerical evaluations of the numerators $\mathcal{N}_i$, as
the denominators have been determined in Section~\ref{codim1} from the
study of varieties of codimension one.  The following discussion is
applicable to any quotient ring $R_n$ with a multiplicity $n$ bigger
than four\footnote{While $R_4$ is an integral domain, it is not a
  unique factorization domain, as in $R_4$ we have
  $\langle12\rangle[12]=\langle34\rangle[34]$. Therefore, codimension-one
  ideals $\langle \mathcal{D}_\alpha \rangle_{R_4}$ may be reducible
  and the common denominator is not unique. Hence, the discussion in
  Section~\ref{codim1}, and in the current section, cannot be directly
  applied in $R_4$.},
thus we will omit the ideal subscripts.

Given a pair of distinct poles $(\mathcal{D}_\alpha ,
\mathcal{D}_\beta)$, we consider the intersection of the associated
varieties, which is equivalent to the variety associated to the sum of
the ideals,
\beq
  V(\big\langle \mathcal{D}_\alpha \big\rangle) \cap V(\big\langle \mathcal{D}_\beta \big\rangle) = V(\big\langle \mathcal{D}_\alpha\big\rangle + \big\langle \mathcal{D}_\beta \big\rangle) = V(\big\langle \mathcal{D}_\alpha, \mathcal{D}_\beta \big\rangle) \, .
\eeq
Contrary to the ideals of codimension one introduced in the previous
section, these ideals $\langle \mathcal{D}_\alpha, \mathcal{D}_\beta
\rangle$ of codimension two do not always correspond to irreducible
varieties. As the numerator may vanish to different orders on the
different branches of $V(\langle \mathcal{D}_\alpha, \mathcal{D}_\beta
\rangle)$, we compute the primary decompositions, as by
Eq.~\eqref{eq:minimal-primary-decomposition}. They read,
\beq
\label{eq:minimal-primary-decomposition-codim2}
\big \langle \mathcal{D}_\alpha , \mathcal{D}_\beta \big \rangle = \bigcap_{l=1}^{n_Q(\langle \mathcal{D}_\alpha , \mathcal{D}_\beta \rangle)} Q_l \, , \\
\eeq
where, by Eq.~\eqref{eq:first-CM-consequence}, we have
\beq
\quad n_Q(\langle \mathcal{D}_\alpha , \mathcal{D}_\beta \rangle) = n_U\big(V(\langle \mathcal{D}_\alpha , \mathcal{D}_\beta \rangle)\big) \, .
\eeq
That is, we are in a special situation where the minimal decomposition
of the associated variety reads
\beq
\label{eq:VarietyDecompositionFromPrimaryDecomposition}
  V(\big \langle \mathcal{D}_\alpha , \mathcal{D}_\beta \big \rangle) = \bigcup_{l=1}^{n_U(V(\langle \mathcal{D}_\alpha , \mathcal{D}_\beta \rangle))} V(Q_l) \, .
\eeq  
For each irreducible variety $V(Q_l)$ we then generate either a single
or a pair of nearby phase-space points, depending on whether the field
$\mathbb{F}$ is taken to be $\mathbb{Q}_p$ or $\mathbb{C}$
respectively, and thus obtain the degree of vanishing of the
numerators $\mathcal{N}_i$. Given a prime ideal $P_l = \big\langle
p_1, \dots, p_r,q_1, \dots, q_4 \big\rangle$, where $J_{\Lambda_7} =
\big\langle q_1, \dots, q_4 \big\rangle$, the phase-space point we
require is a set $(\eta^{(\epsilon)}, \tilde\eta^{(\epsilon)}) \in
\mathbb{F}^{28}$ such that,
\beq
\label{eq:symmetric-approach-point}
p_i(\eta^{(\epsilon)}, \tilde\eta^{(\epsilon)}) = \epsilon^{\kappa_i} \quad \text{and} \quad
q_j(\eta^{(\epsilon)}, \tilde\eta^{(\epsilon)}) = \epsilon^{\kappa_q} \, ,
\eeq
where $\kappa_q$ is the working precision ($\kappa_q \gg \kappa_i$),
and $\kappa_i$ is the largest integer such that $p_i \in
P_l^{\langle\kappa_i\rangle}$. Usually, but not always, one has
$\kappa_i=1$. Such a phase-space point can be built with the method
described in ref.~\cite[Section 3]{DeLaurentis:2022otd}, which is also
easily adapted to $\mathbb{F} = \mathbb{C}$. The parameter $\epsilon$
denotes a small quantity with respect to the absolute value of the chosen
field. For each variety $V(Q_l)$ this results in a constraint on the
numerator of the form
\beq
\mathcal{N}_i(\eta^{(\epsilon)}, \tilde\eta^{(\epsilon)}) \sim \epsilon^{\kappa_l} \;\; \Longrightarrow \;\; \mathcal{N}_i \in P_l^{\langle \kappa_l \rangle} \; .
\eeq

To achieve compact representations of the coefficients, as well as to
be able to reconstruct one pole residue at a time, we wish to interpret
as many as possible of these constraints on the numerators in terms of
a partial fraction decomposition of $\mathcal{C}_i$. To illustrate this let
us now restrict our discussion to those codimension-two ideals $\big
\langle \mathcal{D}_\alpha , \mathcal{D}_\beta \big \rangle$ which are
radical, that is to those ideals such that $Q_l = P_l$ for all
$l$. The extension of the following reasoning to the generic case is
addressed in appendix
\ref{sec:AsymmetricApproachesAndRingExtensions}. Let $\kappa$ be the
largest power such that $\mathcal{N}_i$ vanishes to order $\kappa$ on
all branches $V(Q_l)$ of $V(\big \langle \mathcal{D}_\alpha ,
\mathcal{D}_\beta \big \rangle)$, we have,
\beq
\label{eq:maximal-codimension-symbolic-power-membership}
  \mathcal{N}_i \in \bigcap_l P_l^{\langle\kappa\rangle} \;\, \Longrightarrow \;\, \mathcal{N}_i \in \big\langle \mathcal{D}_\alpha , \mathcal{D}_\beta \big\rangle ^{\langle \kappa \rangle} \, , \quad \text{if} \;\, P_l = Q_l \, .
\eeq
By Eq.~\eqref{eq:second-CM-consequence} we can then explicitly expand the symbolic power as,  
\beq
  \big\langle \mathcal{D}_\alpha , \mathcal{D}_\beta \big\rangle ^{\langle \kappa \rangle} = \big\langle \mathcal{D}_\alpha , \mathcal{D}_\beta \big\rangle ^\kappa =
  \big\langle \mathcal{D}_\alpha^\kappa\,, \;\mathcal{D}_\alpha^{\kappa-1}\mathcal{D}_\beta \,, \; \dots \,, \; \mathcal{D}_\alpha\mathcal{D}_\beta^{\kappa-1} \, , \; \mathcal{D}_\beta^\kappa \big\rangle \, .
\eeq
This can be interpreted in terms of a partial-fraction decomposition
of the rational function,
\beq\label{eq:partial-fraction-decomposition}
  \mathcal{C}_i(\lambda, \tilde\lambda) = \frac{1}{\prod_{j\neq\alpha,\beta} \mathcal{D}_j(\lambda, \tilde\lambda)^{q_{ij}}} \sum_{k=0}^{\kappa} \frac{\mathcal{N}_{ik}(\lambda, \tilde\lambda)}{\mathcal{D}_\alpha^{q_\alpha-\kappa+k}\mathcal{D}_\beta^{q_\beta-k}} \, ,
\eeq  
which is manifestly free from spurious poles. This decomposition is
maximal in the sense that it is not possible to replace $\kappa$ with
$\kappa + 1$. However, it may still be possible to further refine the
decomposition as some $\mathcal{N}_{ik}$ may be proportional to
$\mathcal{D}_\alpha$ or $\mathcal{D}_\beta$. How to obtain and
interpret these additional constraints is addressed in appendix
\ref{sec:AsymmetricApproachesAndRingExtensions}. We perform this
analysis for the vast majority of pairs of poles, ($\mathcal{D}_\alpha$,$\mathcal{D}_\beta$).
The required primary decompositions are presented in Section~\ref{sec:primary-decompositions}.
In some rare cases, the partial
fraction decomposition is expected from the structure of unitarity
cuts, and this analysis is thus not needed. For example, one of the
two bubble coefficients in Section~\ref{eq:bubble_coefficients}
involves two three-mass Gram-determinant poles, but these are clearly
associated to different triple cuts and thus separable.

Since the degree of vanishing of $\mathcal{N}_i$ need not be uniform
on all branches of the ideal, it follows that not all constraints can
be interpreted this way. The extra degree of vanishing of the
numerator beyond $\kappa$ on a particular branch is then purely a
statement about the structure of the numerators
$\mathcal{N}_{ik}$. For example, we observe that the integral
coefficients presented in Section~\ref{Integral_Coefficients} often
diverge less strongly on one of the two branches of
$V\left(\left\langle ⟨7|\Gamma_{34|56}|7\rangle, [7|\Gamma_{34|56}|7]
\right\rangle\right)$, which are given in
Eq.~\eqref{eq:primary_decomposition_zaa22_zbb22}. Therefore, besides a
partial-fraction decomposition, we obtain further information on the
numerators. For instance, if the degree of divergence is lower on the
second branch than on the first one, the numerators corresponding to
the leading poles of $⟨7|\Gamma_{34|56}|7⟩$ and $[7|\Gamma_{34|56}|7]$
may contain contractions of the third generator
$\tilde\Gamma_{12|34|56}$ given in the second primary ideal of Eq.~\eqref{eq:primary_decomposition_zaa22_zbb22}.
Thus, the primary decompositions allow one
to uncover numerator structures with specific vanishing properties
in certain regions of phase space.

So far we have achieved a systematic partial fraction decomposition
between pairs of poles, and a way to identify new spinor structures
for the numerators.  The next step is to combine all the
decompositions and numerator constraints together, so that the
resulting expression is simple.  However, note that in general not all
ideal-membership constraints to which the common numerator
$\mathcal{N}_i$ is subject carry over to the individual
$\mathcal{N}_{ik}$. For example, if $\mathcal{N}_i \in J_1 \cap J_2$,
with $J_1 = \langle \mathcal{D}_\alpha , \mathcal{D}_\beta \rangle^{\langle \kappa_1 \rangle}$
and $J_2 = \langle \mathcal{D}_\gamma , \mathcal{D}_\delta \rangle^{\langle \kappa_2 \rangle}$,
then we could attempt a partial fraction decomposition of the form
\beq\label{eq:partial-fraction-decomposition-chained}
  \mathcal{C}_i(\lambda, \tilde\lambda) = \frac{1}{\prod_{j\neq\alpha,\beta,\gamma,\delta} \mathcal{D}_j(\lambda, \tilde\lambda)^{q_{ij}}} \sum_{k_1=0}^{\kappa_1}\sum_{k_2=0}^{\kappa_2} \frac{\mathcal{N}_{ik_1k_2}(\lambda, \tilde\lambda)}{\mathcal{D}_\alpha^{q_\alpha-\kappa_1+k_1}\mathcal{D}_\beta^{q_\beta-k_1}\mathcal{D}_\gamma^{q_\gamma-\kappa_2+k_2}\mathcal{D}_\delta^{q_\delta-k_2}} \, .
\eeq  
However, this may or may not be a valid decomposition of the
coefficient $\mathcal{C}_i$. To see this, let us consider the
simplified case $\kappa_1=\kappa_2=1$. We can refer back to
Eq.~\eqref{eq:partial-fraction-decomposition} and use the first $J_1$
constraint, that is, we can write $\mathcal{N}_i =
\mathcal{N}_{i0}\mathcal{D}_\alpha +
\mathcal{N}_{i1}\mathcal{D}_\beta$. To then achieve the decomposition
of Eq.~\eqref{eq:partial-fraction-decomposition} we need
$\mathcal{N}_{i0} \in J_2$ and $\mathcal{N}_{i1} \in J_2$. Yet, we are
guaranteed that this is true only if $J_1 \cap J_2 = J_1 \cdot J_2$,
which is not the case in general. On the other hand, failure of the
intersection to equal the product does not automatically imply that an
ansatz of the form of Eq.~\eqref{eq:partial-fraction-decomposition}
has to fail. This makes it highly non-trivial to make use of multiple
constraints while performing a partial-fraction decomposition. In
fact, it is not hard to find cases where, given a certain
partial-fraction decomposition, it becomes impossible to make all
other numerator constraints manifest. This can lead to spurious
singular behaviour even in the absence of spurious poles. That is,
spurious divergences on codimension-two varieties are possible even in
the absence of spurious divergences on codimension-one varieties. A
well known example of this type of spurious singularities is the
appearance of $s_{ij}$ poles in lieu of $\langle ij \rangle$ and/or
$[ij]$ poles in gauge-theory amplitudes. An example of a partial
fraction decomposition which makes it impossible to manifest all
numerator constraints is given in
Section~\ref{partial-fraction-example}.
Because of this subtlety, which warrants further investigation in the
future, for the time being we take an heuristic guess-and-check
approach when combining multiple constraints. That is, we assume that
multiple constraints can be naively combined, for example as in
Eq.~\eqref{eq:partial-fraction-decomposition-chained}, and attempt to
fit the free coefficients in the ansatz for the numerators with the
in-limit reconstruction strategy of
ref.~\cite{DeLaurentis:2019phz}. If this fails, we relax one or more
constraints until the reconstruction succeeds. 

\subsubsection{Codimension-two primary decompositions}\label{sec:primary-decompositions}

In this sub-section we present several primary decompositions which
were used in the present computation. Unless otherwise stated, all
ideals are understood to be taken in the quotient ring $R_7$. The
following is not meant to be a complete list of primary decompositions
for codimension-two ideals at seven point. In fact, including spinor
strings of the form $\langle a | \Gamma_{bc}| d ]$ and three-particle
Mandelstam invariants $s_{abc}$ one obtains hundreds of distinct
codimension-two varieties, which is beyond the current scope. The
notation employed is as in
Eq.~\eqref{eq:minimal-primary-decomposition-codim2}, that is, the
left-hand side consists of a reducible ideal and the right-hand side
gives its decomposition as an intersection of primary ideals. Whenever
a primary ideal $Q_l$ does not correspond to its associated prime
$P_l$, i.e.~if $Q_l\neq\sqrt{Q_l}$, then we also write an equation of
the form $\sqrt{Q_l}=P_l$, with $P_l$ explicitly given by a set of generators.

There are five independent ideals generated by pairs of two-particle
invariants. Only one is not prime, and it splits into a ``collinear''
branch and a ``soft'' branch,
\beq\label{eq:codim2-two-two-particle-invariants}
  \begin{gathered}
    \big\langle ⟨12⟩, ⟨13⟩ \big\rangle = \big\langle ⟨12⟩, ⟨13⟩, ⟨23⟩ \big\rangle \cap \big\langle |1⟩ \big\rangle \, , \\
    \text{while} \quad
    \big\langle ⟨12⟩, ⟨34⟩ \big\rangle \, , \;
    \big\langle ⟨12⟩, [12] \big\rangle \, , \;
    \big\langle ⟨12⟩, [13] \big\rangle \, , \;
    \big\langle ⟨12⟩, [34] \big\rangle \quad \text{are prime}.
  \end{gathered}
\eeq
We note that compared to the codimension-two five-point primary
decompositions presented in ref.~\cite{DeLaurentis:2022otd}, the
ideal generated by all angle brackets is absent. This is explained by
the following observation,
\beq
  \text{codim}\left(\big\langle \langle ij\rangle \;\, \forall \;\, i\neq j \in (1, \dots, n) \; \big\rangle_{R_n}\right) = n-3 \, ,
\eeq
which can be easily checked up to very high multiplicity.

There are also five independent ideals generated by a two-particle
invariant and a parity-invariant spinor string,
\beq\label{eq:codim2-two-particle-invariant-with-sum-two-mandelstams}
  \begin{gathered}
    \big\langle ⟨12⟩, ⟨1|\Gamma_{23}|1] \big\rangle \; = \; \big\langle ⟨12⟩, ⟨13⟩, ⟨23⟩ \big\rangle \cap \big\langle |1⟩ \big\rangle \cap \big\langle ⟨12⟩, [13] \big\rangle \, , \\
    \big\langle ⟨12⟩, ⟨1|\Gamma_{34}|1] \big\rangle \; = \; \big\langle |1⟩ \big\rangle \cap \big\langle ⟨12⟩, ⟨1|\Gamma_{34}|1], ⟨2|\Gamma_{34}|1] \big\rangle \, ,\\
    \big\langle ⟨12⟩, ⟨3|\Gamma_{12}|3] \big\rangle \; = \; \big\langle ⟨12⟩, ⟨13⟩, ⟨23⟩  \big\rangle \cap \big\langle ⟨12⟩, \Gamma_{12}|3] \big\rangle \, ,\\ 
    \text{while} \quad \big\langle ⟨12⟩, ⟨3|\Gamma_{14}|3] \big\rangle \, , \; \big\langle ⟨12⟩, ⟨3|\Gamma_{45}|3] \big\rangle \quad \text{are prime} \, .
  \end{gathered}
\eeq
As the primary decompositions of
Eq.~\eqref{eq:codim2-two-two-particle-invariants} and
Eq.~\eqref{eq:codim2-two-particle-invariant-with-sum-two-mandelstams}
are the same as those in $R_6$, it seems natural to conjecture that
they should in fact hold for all $R_{n\geq6}$.

Proceeding to ideals involving a two-particle invariant and a longer
spinor chain we identify six ideals, of which three are not primary,
\beq\label{eq:codim2-two-particle-invariant-with-longer-spinor-string}
  \begin{gathered}
    \big\langle ⟨12⟩, ⟨7|\Gamma_{34|56}|7⟩ \big\rangle = \big\langle ⟨12⟩, ⟨17⟩, ⟨27⟩ \big\rangle \cap \big\langle ⟨12⟩, \Gamma_{12|34}|7⟩ \big\rangle \, , \\
      \big\langle [12], ⟨7|\Gamma_{34|56}|7⟩ \big\rangle = \big\langle [12], \Gamma_{12}|7⟩ \big\rangle \cap \big\langle [12], ⟨7|\Gamma_{56}|1], ⟨7|\Gamma_{56}|2] \big\rangle \, , \\
        \big\langle ⟨17⟩, ⟨7|\Gamma_{34|56}|7⟩ \big\rangle = \big\langle ⟨17⟩, |7⟩⟨7| \big\rangle \cap \big\langle ⟨12⟩, ⟨17⟩, ⟨27⟩ \big\rangle \cap \big\langle ⟨17⟩, ⟨7|\Gamma_{56}|2], ⟨1|\Gamma_{56}|2] \big\rangle \, , \\
           \text{while} \quad \big\langle [17], ⟨7|\Gamma_{34|56}|7⟩ \big\rangle \, , \; \big\langle ⟨13⟩, ⟨7|\Gamma_{34|56}|7⟩ \big\rangle \, ,\; \big\langle [13], ⟨7|\Gamma_{34|56}|7⟩ \big\rangle \quad \text{are prime} \, .
  \end{gathered}
\eeq
In the above $\big\langle ⟨17⟩, |7⟩⟨7| \big\rangle$ is clearly not
radical, as it contains the outer product of $|7⟩$ with
itself, i.e.~it is primary but not prime. The associated
prime is simply the soft ideal,
\beq\label{eq:non-radical-ideal-oneseven}
  \sqrt{\big\langle ⟨17⟩, |7⟩⟨7| \big\rangle} = \big\langle |7⟩ \big\rangle \, .
\eeq

We also make extensive use of two primary decompositions of
codimension-two ideals generated by pairs of spinor chains,
\begin{align}
   \big\langle ⟨7|\Gamma_{12}|7], ⟨7|\Gamma_{34|56}|7⟩ \big\rangle &=  \big\langle ⟨12⟩, ⟨17⟩, ⟨27⟩ \big\rangle \; \cap \; \big\langle [12], \Gamma_{12}|7⟩ \big\rangle \; \cap \; \big\langle ⟨7|\Gamma_{12}|7], |7⟩⟨7| \big\rangle \nonumber \\
&\phantom{=} \quad \cap  \; \big\langle ⟨7|\Gamma_{34}|7], ⟨7|\Gamma_{56}|7], ⟨7|\Gamma_{34|56}|7⟩, [7|\Gamma_{34|56}|7] \big\rangle  \, , \\
    \big\langle ⟨7|\Gamma_{34|56}|7⟩, [7|\Gamma_{34|56}|7] \big\rangle &= \big\langle ⟨7|\Gamma_{34}|7], ⟨7|\Gamma_{56}|7], ⟨7|\Gamma_{34|56}|7⟩, [7|\Gamma_{34|56}|7] \big\rangle  \nonumber \\
&\phantom{=} \quad \cap  \; \big\langle  ⟨7|\Gamma_{34|56}|7⟩, [7|\Gamma_{34|56}|7], \tilde\Gamma_{12|34|56} \big\rangle \, .
\label{eq:primary_decomposition_zaa22_zbb22}
\end{align}
In particular, the latter provides significant constraints, both in
terms of partial fractions and in terms of numerator spinor
structures, since the involved polynomials are of high degree. We
recall that the tilde denotes anti-symmetrization, as defined in
Eq.~\eqref{eq:gamma_tilde}. One of the primaries is again not radical,
and it has the same associated prime as
Eq.~\eqref{eq:non-radical-ideal-oneseven},
\beq \sqrt{\big\langle
⟨7|\Gamma_{12}|7], |7⟩⟨7| \big\rangle} = \big\langle |7⟩ \big\rangle \, .
\eeq

Lastly, we consider a few codimension two ideals involving three-mass
Gram determinants. These are often not radical, but generally they are
primary,
\beq\label{eq:non-radical-delta}
  \begin{aligned}
    \big\langle ⟨12⟩, Δ_3(1,2,3,4) \big\rangle &= \big\langle ⟨12⟩,  (s_{567}-s_{34})^2 \big\rangle \, , \\
    \big\langle ⟨1|\Gamma_{34}|2], Δ_3(1,2,3,4) \big\rangle &= \big\langle ⟨1|\Gamma_{34}|2], (s_{134}-s_{234})^2 \big\rangle \, ,
  \end{aligned}
\eeq
where we have made explicit the existence of a perfect-square
polynomial in the ideals. Then, the radicals can be shown to be,
\beq\label{eq:radical-delta}
  \begin{aligned}
    \sqrt{\big\langle ⟨12⟩, Δ_3(1,2,3,4) \big\rangle} &= \big\langle ⟨12⟩,  (s_{567}-s_{34}) \big\rangle \, , \\
    \sqrt{\big\langle ⟨1|\Gamma_{34}|2], Δ_3(1,2,3,4) \big\rangle} &= \big\langle ⟨1|\Gamma_{34}|2], (s_{134}-s_{234}) \big\rangle \, .
  \end{aligned}
\eeq
The attentive reader may recognize that $⟨1|\Gamma_{34}|2]$ is not one
of the poles listed in Eq.~\eqref{eq:poles-list}, however it is often a zero of the
residue of the three-mass Gram pole, hence we choose to include
it here. Finally, there is also strong evidence for the following
primary decomposition
\beq\label{eq:last-prime-dec}
  \begin{gathered}
    \big\langle ⟨7|\Gamma_{34|56}|7⟩, Δ_3(3,4,5,6) \big\rangle = \big\langle ⟨7|\Gamma_{34|56}|7⟩,  Δ_3(3,4,5,6), \tilde\Gamma_{34|56}|7⟩⟨7|\tilde\Gamma_{34|56} \big\rangle \, , \\
    \sqrt{\big\langle ⟨7|\Gamma_{34|56}|7⟩, Δ_3(3,4,5,6) \big\rangle} = \big\langle Δ_3(3,4,5,6), \tilde\Gamma_{34|56}|7⟩ \big\rangle \, .
  \end{gathered}
\eeq
The primality of all prime ideals can be proven via the test presented
in Appendix B.3 of ref.~\cite{DeLaurentis:2022otd}, except the very
last one of Eq.~\eqref{eq:last-prime-dec}, which remains to be
proven. It is also possible to explicitly check that the intersection
of the primaries equals the reducible ideal, and whenever a primary is
not radical one can check that $\sqrt{Q_l} = P_l$ by verifying
$\text{dim}(Q_l) = \text{dim}(P_l)$ and that $Q_l / P_l^\infty =
\langle 1 \rangle$, where the latter operation denotes ideal
saturation. In the ancillary files we provide a \texttt{Python} script that performs these checks.

\subsubsection{Example of simultaneous constraints and spurious singularities}
\label{partial-fraction-example}

Let us now consider an example to illustrate the subtetly with
combining multiple numerator constraints when interpreting some in
terms of partial-fraction decompositions. We take a part of the
integral coefficient $\tilde c^{(2)}_{12\times56}$ from
Eq.~\eqref{ct12x56m2}, namely,
\beq
  \begin{gathered}
    \mathcal{C} = \frac{[2|\tilde\Gamma_{12|34|56}|1⟩}{s_{12}⟨7|\Gamma_{34|56}|7⟩[7|\Gamma_{34|56}|7]} \, .
  \end{gathered}
\eeq
The numerator belongs to the following two ideals: $J_1 = \big\langle
⟨7|\Gamma_{34|56}|7⟩, [7|\Gamma_{34|56}|7], \tilde\Gamma_{12|34|56}
\big\rangle$ and $J_2 = \big\langle ⟨12⟩, [12] \big\rangle$. That is,
the numerator $\mathcal{N}$ belongs to $J_1 \cap J_2$.
One can recognize $J_1$ as the second primary ideal in the primary
decomposition of Eq.~\eqref{eq:primary_decomposition_zaa22_zbb22}. At
the same time, $\mathcal{N}$ does not belong to the first primary
ideal in that same decomposition. Therefore, the two poles
$⟨7|\Gamma_{34|56}|7⟩$ and $[7|\Gamma_{34|56}|7]$ cannot be separated
without introducing spurious poles\footnote{Note that if one were to
  introduce a spurious pole of, say, the form $\spab7.34.7$, then the
  numerator would indeed vanish on both branches and a
  partial-fraction decomposition would be possible.}. However, the
numerator does belong to the ideal $\big\langle ⟨12⟩, [12]
\big\rangle$, even if this is perhaps not entirely manifest. It
becomes apparent by expanding the above as,
\beq
  \begin{gathered}
    [2|\tilde\Gamma_{12|34|56}|1⟩ = ⟨12⟩[2|\Gamma_{56|34}|2]+[12]⟨1|\Gamma_{56|34}|1⟩ \, ,
  \end{gathered}
\eeq
which can then be written as a partial fraction decomposition of the form,
\beq
\label{eq:partial-fractioned-example}
\mathcal{C} = \frac{[2|\Gamma_{34|56}|2]}{[12]⟨7|\Gamma_{34|56}|7⟩[7|\Gamma_{34|56}|7]} + \frac{⟨1|\Gamma_{34|56}|1⟩}{⟨12⟩⟨7|\Gamma_{34|56}|7⟩[7|\Gamma_{34|56}|7]} \, .
\eeq
Note that now it has become impossible to have a
$\tilde\Gamma_{12|34|56}$ factor in either numerator, as their mass
dimensions are not sufficient. We can see this might happen as product
and intersection are not equal for the ideals $J_1$ and $J_2$.
These kinds of consideration have implications on the stability of the
expressions. For instance, the latter partial-fraction form of
Eq.~\eqref{eq:partial-fractioned-example} is potentially unstable near
the variety associated to the second primary ideal in the primary
decomposition of Eq.~\eqref{eq:primary_decomposition_zaa22_zbb22}
(double-pole cancelling to give a simple pole in the sum of the
fractions).

\section{Integral Coefficients}
\label{Integral_Coefficients}
\begin{figure}
\begin{center}
\includegraphics[width=0.44\textwidth,angle=270]{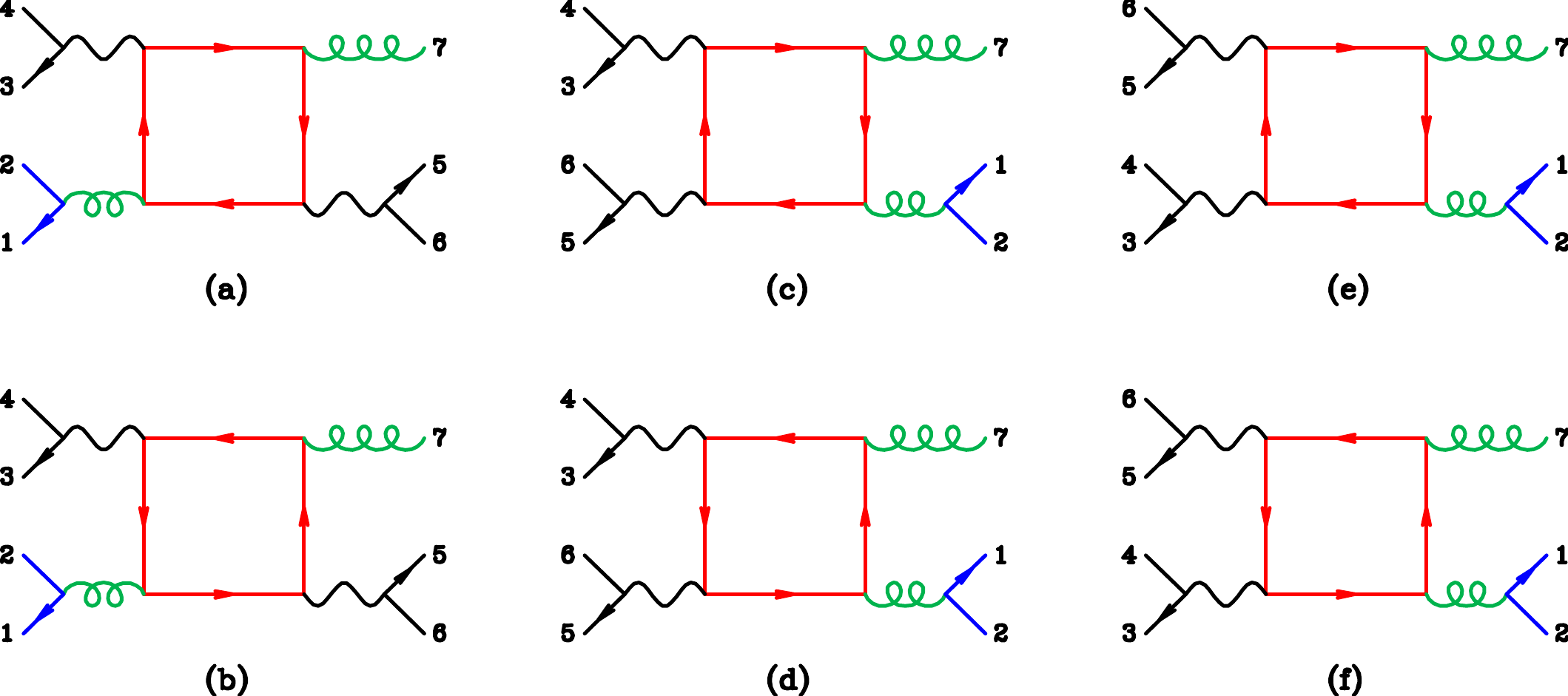}
\caption{Box diagrams containing a quark loop\label{boxdiags}}
\end{center}
\end{figure}

In this section we present results for the one-loop master-integral
coefficients for the process under consideration, starting for the six
one-loop diagrams shown in Fig.~\ref{boxdiags} which contribute to
doubly-resonant $Z$-boson pair production in association with a
jet. These results have been calculated using standard analytic
techniques~\cite{Britto:2004nc,Forde:2007mi,Badger:2008cm,Mastrolia:2009dr,Ellis:2011cr}
and subsequently simplified using the methods of
Section~\ref{Geometric}.
The one-loop colour amplitudes have the decomposition,
\beq
{\cal A}_7^{\oneloop,B}(1,2,3,4,5,6,7)=
    4 i g_s e^4 \frac{g_s^2}{16 \pi^2}  (t^B)_{{i_1 \ib_2}} A_7^{\oneloop}(1,2,3,4,5,6,7),
\eeq
where the $SU(3)$ colour matrix in the fundamental representation, $t^B$, is normalized such that ${\rm tr}~t^A t^B = \delta^{AB}$.
The quark and antiquark colour indices are $i_1$ and $\ib_2$. The colour stripped amplitude can be expressed
in terms of scalar integrals and rational terms $r$.
\begin{eqnarray} \label{scalarreduction}
  &&A_7^{\oneloop}(1^{h_1},2^{h_2},3^{h_3},4^{h_4},5^{h_5},6^{h_6},7^{h_7})
  =  \frac{\bar\mu^{4-n}}{r_\Gamma}\frac{1}{i \pi^{n/2}} \int {\rm d}^n \ell
 \, \frac{{\rm Num}(\ell)}{\prod_i d_i(\ell)} \nonumber \\
 &=&
 \sum_{i,j,k} {d}_{\{i\sep j\sep k\}}(1^{h_1},2^{h_2},3^{h_3},4^{h_4},5^{h_5},6^{h_6},7^{h_7}) \, D_0(p_i, p_j, p_k ;m)  \nonumber \\
&+& \sum_{i,j} {c}_{\{i\sep j\}}(1^{h_1},2^{h_2},3^{h_3},4^{h_4},5^{h_5},6^{h_6},7^{h_7}) \,  C_0(p_i,p_j ;m)   \nonumber \\
 &+& \sum_{i} {b}_{\{i\}}(1^{h_1},2^{h_2},3^{h_3},4^{h_4},5^{h_5},6^{h_6},7^{h_7}) \, B_0(p_i;m) \nonumber \\
 &+& {r}(1^{h_1},2^{h_2},3^{h_3},4^{h_4},5^{h_5},6^{h_6},7^{h_7})\, .
\end{eqnarray}
The definitions of the scalar integrals are given in Appendix~\ref{Integrals}.
As in Eq.~\eqref{momentumlabelling} we present results for one specific helicity choice,
\beq
A_7^{\oneloop}(1^-,2^+,3^-,4^+,5^-,6^+,7^+).
\eeq
Other helicities are obtained by permutation of the arguments.

The tree amplitude that we presented in Eqs.~(\ref{colourstrip},\ref{bremsampA}) was exactly the amplitude 
for production of a gluon and two lepton pairs off a massless, unit-charge, quark line.
The production of the two lepton pairs was mediated by a virtual photon, and the propagators and 
couplings appropriate for $Z$-boson exchange were added subsequently 
as detailed in Eq.~\eqref{tree_amplitude_with_couplings}. For the quark loop amplitudes, because of the mass 
running in the quark loop, we need to maintain helicity information. Therefore the amplitudes that we present below
generalize the coupling of the ``photon'' to the massive quark line as follows,
\beq
-i e \gamma^\mu \to -i e \big(v_R \gamma^\mu \gamma_R+ v_L \gamma^\mu \gamma_L\big) \, , \;\, \text{where} \;\; \gamma_{R/L}=\frac{1}{2} \big(1\pm\gamma_5\big)\,.
\eeq
The amplitude for true photon exchange is recovered by setting $v_L=v_R=1$.
Given this helicity information it is straightforward to add the couplings and propagators appropriate for $ZZ$ production.

In several cases we observe that, while overall a given coefficient
$\mathcal{C}$ may be of mixed symmetry under a certain swap operation
$\mathcal{S}$, some of its poles are actually either fully symmetric or
anti-symmetric. In these cases, we find it convenient to split the
coefficients in their symmetric and anti-symmetric parts
with respect to~$\mathcal{S}$ as,
\beq\label{eq:symmetry-decomposition}
\mathcal{C}^{(S)} = \frac{1}{2} \Big[\mathcal{C}+\mathcal{C}\big|_{\mathcal{S}}\Big] \; , \quad
\mathcal{C}^{(A)} = \frac{1}{2} \Big[\mathcal{C}-\mathcal{C}\big|_{\mathcal{S}}\Big] \; .
\eeq

\subsection{Results for box coefficients}
We begin with a decomposition of the box coefficients in terms of the mass $m$ and the
left-handed and right-handed couplings of the vector bosons to the
quark loop
\begin{eqnarray} \label{basicexpansion}
d_{\{i\sep j\sep k\}}&=\;(v_L^2+v_R^2)\, &\big[m^0 d_{\{i\sep j\sep k\}}^{(0)}+m^2 d_{\{i\sep j\sep k\}}^{(2)}+m^4 d_{\{i\sep j\sep k\}}^{(4)}\big] \nonumber \\
 &\;+\;v_L v_R &\big[ m^2 \tilde{d}_{\{i\sep j\sep k\}}^{(2)}+m^4 \tilde{d}_{\{i\sep j\sep k\}}^{(4)} \big] \, .
\end{eqnarray}
$i,j$ and $k$ thus represent the outgoing momenta at three of the four corners of the box.
Analogous expansions will follow for triangle and bubble coefficients. 
\subsubsection{Results for $\{12\sep 34\sep 56\}$ box}
The result for the quark mass-independent piece of this
box is,
\begin{eqnarray}
\label{d12x34x560}
d_{\{12\sep34\sep56\}}^{(0)}&=&\Bigg\{-
\frac{\spa5.7^2\*\spab7.12.4^2\*\spaba7.56.34.1^2
\*(s_{127}\*s_{567}-s_{12}\*s_{56})}
{4\*\spa5.6\*\spa1.2\*\spb3.4\*\spaba7.56.12.7^4}\Bigg\}\nonumber\\
&+&\Bigg\{\;\Bigg\}_{1 \leftrightarrow 5,2 \leftrightarrow 6}
\end{eqnarray}

It is manifestly symmetric under a swap of the opposite corners of the
box diagram, that is $(1,2) \leftrightarrow (5,6)$. To present the
result for the $m^2$ coefficient we employ the decomposition of
Eq.~\eqref{eq:symmetry-decomposition}, as the $\langle
7|\Gamma_{34|56}|7\rangle^3$ and the $[7|\Gamma_{34|56}|7]^2$ poles have simple properties
under the interchange. We also remind the reader that there exist identities such as
\beq
\spaba7.34.56.7=-\spaba7.12.56.7\, .
\eeq
The symmetric part reads
\begin{eqnarray}
\label{12x34x56explicit}
&&d_{\{12\sep34\sep56\}}^{(2,S)} =\Bigg\{
\frac{\spb6.7^2\*\spab3.12.7\*\spbab2.34.56.7
\*\tspbaba2.12.34.56.3}
{4\*\spb1.2\*\spa3.4\*\spb5.6\*\spaba7.34.56.7
\*\spbab7.34.56.7^2}
\nonumber\\&&
-\frac{\spa5.7^2\*\spab7.12.4\*\spaba1.34.56.7
\*\tspbaba4.12.34.56.1}
{\spa1.2\*\spb3.4\*\spa5.6\*\spaba7.34.56.7^3}
-\frac{\spa1.7\*\spb6.7^2\*\spab7.56.4\*\tspbaba2.12.34.56.3}
{4\*\spb5.6\*\spaba7.34.56.7^2
\*\spbab7.34.56.7}
\nonumber\\&&
+\frac{\spa1.3\*\spb2.7\*\spb6.7^2\*\spab3.12.7}
{8\*s_{12}\*\spa3.4\*\spb5.6\*\spbab7.34.56.7}
+\frac{\spa1.3\*\spb4.6\*\spa5.7\*\spaba1.34.56.7
\*\spabab7.12.34.56.7}
{4\*\spa1.2\*s_{34}\*s_{56}\*\spaba7.34.56.7^2}
\nonumber\\&&
+\frac{\spa1.5\*\spb4.6\*\spab7.56.4\*\spab7.12.7
\*\spaba1.34.56.7}
{4\*\spa1.2\*\spb3.4\*s_{56}\*\spaba7.34.56.7^2}
-\frac{\spa1.3\*\spa3.5\*\spb6.7\*\spaba1.34.56.7}
{8\*\spa1.2\*\spa3.4\*\spa5.6\*\spb5.6
\*\spaba7.34.56.7}
\nonumber\\&&
-\frac{\spa1.5\*\spb4.6\*\spb4.7\*\spaba1.34.56.7}
{8\*\spa1.2\*\spb3.4\*\spa5.6\*\spb5.6
\*\spaba7.34.56.7}
+\frac{\spa1.5\*\spb2.7\*\spb4.6\*\spab7.56.4}
{8\*\spb3.4\*\spa5.6\*\spb5.6\*\spaba7.34.56.7}
\nonumber\\&&
-\frac{\spb2.6\*\spb2.7\*\spa3.7^2\*\spb6.7}
{8\*\spb1.2\*\spa3.4\*\spb5.6\*\spaba7.34.56.7}
-\frac{\spa3.5^2\*\spaba1.34.56.7^2}
{4\*\spa1.2\*\spa3.4\*\spa5.6\*\spaba7.34.56.7^2}
\nonumber\\&&
-\frac{\spb4.6^2\*\spaba1.34.56.7^2}
{2\*\spa1.2\*\spb3.4\*\spb5.6
\*\spaba7.34.56.7^2}
+\frac{\spb2.4\*\spa5.7\*\spab5.36.4
\*\spaba1.34.56.7}
{2\*\spb3.4\*\spa5.6
\*\spaba7.34.56.7^2}
\nonumber\\&&
-\frac{\spa3.5\*\spa1.3\*\spab7.12.6
\*\spaba1.34.56.7}
{2\*\spa1.2\*\spa3.4\*\spaba7.34.56.7^2}
-\frac{\spb2.6\*\spa3.7^2\*\spb6.7\*\spaba1.34.56.7}
{4\*\spa3.4\*\spb5.6\*\spaba7.34.56.7^2}
\nonumber\\&&
+\spa3.5\*\spaba1.34.56.7
\*\frac{\spa1.7\*\spb3.4\*\spa3.5-\spa1.5\*\spab7.12.4}
{4\*\spa1.2\*\spa5.6\*\spaba7.34.56.7^2}
+\frac{\spa1.7\*\spa3.4\*\spb4.6^2\*\spaba1.34.56.7}
{2\*\spa1.2\*\spb5.6\*\spaba7.34.56.7^2}
\nonumber\\&&
+\frac{\spa1.5\*\spb4.6\*\spab7.12.4
\*\spaba1.34.56.7}
{2\*\spa1.2\*\spb3.4\*\spaba7.34.56.7^2}
-\frac{\spb1.2\*\spa1.7\*\spb4.6^2\*\spaba1.34.56.7}
{2\*\spb3.4\*\spb5.6\*\spaba7.34.56.7^2}
-\frac{\spa1.3\*\spa1.5\*\spb4.6}
{\spa1.2\*\spaba7.34.56.7}
\nonumber\\&&
-3\*\frac{\spa1.3\*\spb2.6\*\spa3.5}
{4\*\spa3.4\*\spaba7.34.56.7}
-\frac{\spa1.3\*(\spa1.3\*\spab5.12.6
-\spa1.3\*\spab5.34.6
+\spa1.5\*\spa3.5\*\spb5.6)}
{2\*\spa1.2\*\spa3.4\*\spaba7.34.56.7}
\nonumber\\&&
-3\*\frac{\spa1.5\*\spa1.3
\*(2\*\spab5.12.4-\spb3.4\*\spa3.5)}
{16\*\spa1.2\*\spa5.6\*\spaba7.34.56.7}
-\frac{\spa1.3\*\spab1.57.6\*\spab3.57.6}
{2\*\spa1.2\*\spa3.4\*\spb5.6\*\spaba7.34.56.7}
\nonumber\\&&
-\spa1.5\*\frac{8\*\spa1.2\*\spb2.4\*\spb4.6\*\spa5.6
+6\*\spb1.4\*\spa1.5\*\spa1.6\*\spb4.6
+3\*\spb1.4\*\spa1.5^2\*\spb4.5
+3\*\spa1.5\*\spb2.4\*\spa2.6\*\spb4.6}
{16\*\spa1.2\*\spb3.4\*\spa5.6\*\spaba7.34.56.7}
\nonumber\\&&
+\frac{\spb4.6^2\*\spab1.34.2}
{2\*\spb3.4\*\spb5.6\*\spaba7.34.56.7}\Bigg\}+\Bigg\{\;\Bigg\}_{1\leftrightarrow 5,2\leftrightarrow 6}
\end{eqnarray}

The anti-symmetric part reads
\begin{eqnarray}
  \label{d123456m2asy}
&&d_{\{12\sep34\sep56\}}^{(2,A)}=\Bigg\{
-\frac{\spa1.7\*\spa3.5\*\spb4.6\*\spaba1.34.56.7
\*\spabab7.12.34.56.7}
{4\*\spa1.2\*s_{34}\*s_{56}\*\spaba7.12.56.7^2}
\nonumber\\&&
+\frac{\spaba5.34.12.7
\*(-2\*\spa1.3\*\spb2.7\*\spab5.12.4
+\spa1.3\*\spb2.7\*\spb3.4\*\spa3.5+
\spa1.5\*\spb4.7\*\spab3.14.2)}
{8\*s_{12}\*s_{34}\*\spa5.6\*\spaba7.12.56.7}
\nonumber\\&&
+\frac{\spa1.5\*\spb2.7\*\spb4.6\*\spab7.56.4}
{8\*\spb3.4\*s_{56}\*\spaba7.12.56.7}
+\frac{\spa1.3\*\spb2.7\*\spb6.7^2\*\spab3.12.7}
{8\*s_{12}\*\spa3.4\*\spb5.6\*\spbab7.12.56.7}\Bigg\}-\Bigg\{\;\Bigg\}_{1\leftrightarrow 5,2\leftrightarrow 6}
\end{eqnarray}

The $m^4$ piece reads
\begin{eqnarray}
\label{d4one}
d_{\{12\sep34\sep56\}}^{(4)}&=&\frac{\tspbaba2.34.12.56.1}
{s_{12}\*s_{34}\*s_{56}
\*\spaba7.34.56.7}\*
\left(\frac{\tspbaba4.34.12.56.3
\*\tspbaba6.34.12.56.5}
{\spaba7.34.56.7\*\spbab7.34.56.7}
-\spa3.5\*\spb4.6
\right)\qquad
\end{eqnarray}

The helicity flip pieces of this box amplitude (c.f.~Eq.~\eqref{basicexpansion}) are given by,
\begin{eqnarray}
  \label{dt123456m2}
&&\tilde{d}_{\{12\sep34\sep56\}}^{(2)}=
\frac{\spa1.7\*\spa3.5\*\spb4.6
\*\spaba1.34.56.7\*\spabab7.12.34.56.7}
{\spa1.2\*s_{34}\*s_{56}\*\spaba7.34.56.7^2}
+\frac{\spa3.5\*\spb2.7^2\*\spb4.6}
{2\*\spb1.2\*s_{34}\*s_{56}}
\nonumber\\&&
-\frac{\spa3.5\*\spb2.6\*\spb2.7
\*\spabab7.12.34.56.4}
{2\*\spb1.2\*s_{34}\*s_{56}\*\spaba7.34.56.7}
+\frac{\spa1.3\*\spa1.5\*\spb4.6
\*\spabab7.12.34.56.7}
{2\*\spa1.2\*s_{34}\*s_{56}\*\spaba7.34.56.7}
\nonumber\\&&
-\frac{\spa3.5\*\spb4.7\*\spab1.27.6
\*\spaba1.34.56.7}
{2\*\spa1.2\*s_{34}\*s_{56}\*\spaba7.34.56.7}
-\frac{\spa3.5\*\spb2.7^2\*\spb6.7\*\spab3.56.7}
{2\*\spb1.2\*\spa3.4\*s_{56}\*\spbab7.34.56.7}
\end{eqnarray}

and
\begin{equation}
\label{dt12x34x56m4}
\tilde{d}_{\{12\sep34\sep56\}}^{(4)}=
\frac{2\*\spa3.5\*\spb4.6\*\tspbaba2.34.12.56.1}
{s_{12}\*s_{34}\*s_{56}\*\spaba7.34.56.7}
\end{equation}

Note that Eqs.~\eqref{d4one} and \eqref{dt12x34x56m4} are symmetric under
the exchanges $3\leftrightarrow 5, 4\leftrightarrow 6$. This will be
important in the following because these functions also supply the
$m^4$ pieces of the box symmetric under this exchange,
i.e.~$d_{\{56\sep12\sep34\}}^{(4)}$ and
$\tilde{d}_{\{56\sep12\sep34\}}^{(4)}$.
\subsubsection{Results for $\{12\sep56\sep34\}$ box}
The box coefficients for $d_{\{12\sep56\sep34\}}$ are all obtained from the above results by exchange,
\beq
\label{d4three}
d_{\{12\sep56\sep34\}}^{(i)}= \left. d_{\{12\sep34\sep56\}}^{(i)} \right|_{3 \leftrightarrow 5, 4 \leftrightarrow 6} ,\;\;\;
\tilde{d}_{\{12\sep56\sep34\}}^{(i)}= \left. \tilde{d}_{\{12\sep34\sep56\}}^{(i)} \right|_{3 \leftrightarrow 5, 4 \leftrightarrow 6} \, .
\eeq
\subsubsection{Results for $\{56\sep12\sep34\}$ box}
All the coefficients for this box are fully symmetric under the exchange ${3 \leftrightarrow 5,4 \leftrightarrow 6}$,
which is a symmetry of the relevant diagrams, Figs.~\ref{boxdiags}(a) and \ref{boxdiags}(b).
The mass-independent piece is determined by Eq.~\eqref{d12x34x560},
\beq
d_{\{56\sep12\sep34\}}^{(0)}= \left. d_{\{12\sep34\sep56\}}^{(0)} \right|_{1 \to 5, 2 \to 6, 3\to 1, 4 \to 2, 5 \to 3, 6 \to 4} \, .
\eeq
The coefficient proportional to $m^2$ has certain elements in
common with the suitably permuted ${d}_{\{12\sep34\sep56\}}^{(2,S)}$ from Eq.~\eqref{12x34x56explicit} so we write,
\begin{eqnarray}
&&d_{\{56\sep12\sep34\}}^{(2)}
-d_{\{12\sep34\sep56\}}^{(2,S)} \big|_{\{1 \to 5,2\to 6,3\to 1,4\to 2,5\to 3,6\to 4\}}\nonumber \\
&=&\Bigg\{\frac{\spb2.4\*\spa5.7
\*\spabab7.34.12.56.7\*\big(\spa1.3\*\spaba5.12.34.7
+2\*\spa1.2\*\spa3.5\*\spab7.34.2\big)}
{4\,\*s_{12}\*s_{34}\*\spa5.6\*\spaba7.34.56.7^2}
\nonumber\\&&
-\frac{\spa3.5\*\spb4.7\*\spab1.34.2
\*\spaba5.12.34.7}
{4\,\*s_{12}\*s_{34}\*\spa5.6\*\spaba7.34.56.7}
-3\*\frac{\spa1.3\*\spb2.7\*\spa3.5\*\spaba5.12.34.7}
{8\,\*s_{12}\*\spa3.4\*\spa5.6\*\spaba7.34.56.7}
\nonumber\\&&
-\frac{\spa1.3\*\spa1.5\*\spb4.7\*\spaba5.12.34.7}
{8\,\*\spa1.2\*s_{34}\*\spa5.6\*\spaba7.34.56.7}
-3\*\frac{\spb2.4\*\spb2.7\*\spa3.5\*\spaba5.12.34.7}
{8\,\*\spb1.2\*s_{34}\*\spa5.6\*\spaba7.34.56.7}
\nonumber\\&&
-3\*\frac{\spb2.4\*\spa3.5\*\spb6.7\*\spab7.34.2}
{8\,\*\spb1.2\*s_{34}\*\spaba7.34.56.7}
+\frac{\spa1.3\*\spa1.5\*\spb4.7\*\spb6.7}
{16\,\*\spa1.2\*s_{34}\*s_{56}}
\nonumber\\&&
+\frac{\spa3.5\*\spb4.7\*\spb6.7\*\spab1.34.7
\*\spab1.56.7}
{16\,\*\spa1.2\*s_{34}\*s_{56}\*\spbab7.34.56.7}\Bigg\} + \Bigg\{\;\Bigg\}_{3\leftrightarrow 5,4\leftrightarrow 6}
\end{eqnarray}

The $m^4$ term is supplied by Eq.~\eqref{d4one},
\beq
\label{d4four}
d_{\{56\sep12\sep34\}}^{(4)}=d_{\{12\sep34\sep56\}}^{(4)}
\eeq
For the box coefficients for terms with a helicity flip on the massive quark line we have, 
\begin{eqnarray}&&
\tilde{d}_{\{56\sep12\sep34\}}^{(2)}=\Bigg\{\frac{1}
{4\*s_{34}\*s_{56}}\*\bigg[\spa3.5\*\spb4.6\*\spabab7.56.12.34.7\*\frac{(2\*\spab7.34.2^2-\spb1.2^2\*\spa1.7^2)}
{\spb1.2\*\spaba7.34.56.7^2}
\nonumber\\&&
+2\*\frac{\spa1.3^2\*\spb3.4\*\spb6.7\*\spaba7.34.12.5}
{\spa1.2\*\spaba7.34.56.7}
-2\*\frac{\spa1.3\*\spa5.6\*\spb4.7\*\spb6.7^2\*\spab1.34.7}
{\spa1.2\*\spbab7.12.56.7}
\nonumber\\&&
+\frac{\spa3.7\*\spb4.7\*\spb6.7\*(\spa1.3\*\spb2.3\*\spa5.7-\spa1.5\*\spab7.15.2)}
{\spaba7.34.56.7}
\nonumber\\&&
-2\*\frac{\spa1.3\*\spb6.7\*(\spb2.6\*\spb3.4\*\spa3.7\*\spa5.6
-\spa5.7\*\spb2.4\*(s_{34}+s_{37}))}
{\spaba7.34.56.7}
\nonumber\\&&
-\frac{\spa3.5\*(\spb2.4\*\spb2.6\*\spabab7.56.12.34.7
+\spb4.7\*\spb6.7\*\spab7.34.2^2)}
{\spb1.2\*\spaba7.34.56.7}\bigg]\Bigg\}+\Bigg\{\;\Bigg\}_{3 \leftrightarrow 5,4 \leftrightarrow 6}
\end{eqnarray}

The $m^4$ term is supplied by Eq.~\eqref{dt12x34x56m4},
\beq
\label{d4two}
\tilde{d}_{\{56\sep12\sep34\}}^{(4)}=\tilde{d}_{\{12\sep34\sep56\}}^{(4)}\, .
\eeq
Because a rank two box integral is cut constructible, the vanishing of the rational piece requires the
following relationship,
\beq
\label{rational_relation}
  \tilde{d}_{\{56\sep12\sep34\}}^{(4)}=\tilde{c}_{\{7\sep12\}}^{(2)}+\tilde{c}_{\{7\sep34\}}^{(2)}+\tilde{c}_{\{7\sep56\}}^{(2)}
                               +\tilde{c}_{\{34\sep56\}}^{(2)}+\tilde{c}_{\{12\sep56\}}^{(2)}+\tilde{c}_{\{12\sep34\}}^{(2)} \, .
\eeq
The $c$ coefficients in this Eq.~\eqref{rational_relation} are presented in subsection~\ref{triangle_coefficients}.     

\subsection{Results for triangle coefficients}
\label{triangle_coefficients}
The triangle coefficients have no terms which are
quartic in the mass of the quark,
\beq
c_{\{i\sep j\}}=(v_L^2+v_R^2)\, \big[m^0 c_{\{i\sep j\}}^{(0)}+m^2 c_{\{i\sep j\}}^{(2)}\big] +v_L v_R \big[ m^2 \tilde{c}_{\{i\sep j\}}^{(2)}\big]\,.
\eeq
\subsubsection{Results for $\{34\sep 56\}$ triangle}
\begin{eqnarray}&&
  c_{\{34\sep56\}}^{(0)}=\Bigg\{
\frac{\spa1.7\*\spa3.7\*\spab7.12.6
\*(s_{35}+s_{36}+s_{45}+s_{46})
\*(-2\*\spa1.5\*\spab7.12.4-\spa1.7\*\spab5.36.4)}
{2\*\spa1.2\*\spaba7.12.56.7^3}
\nonumber\\&&
+\frac{\spa1.7\*\spa3.7^2\*\spab7.12.6^2\*\Delta_3(3,4,5,6)
\*(2\*\spa1.7\*(s_{35}+s_{36}+s_{45}+s_{46})
-\spaba1.34.56.7)}
{4\*\spa1.2\*\spa3.4\*\spb5.6\*\spaba7.12.56.7^4}
\nonumber\\&&
+\frac{\spa1.7\*\spa3.7\*\spab7.12.6
\*(s_{35}+s_{36}+s_{45}+s_{46})^2
\*(2\*\spa1.7\*\spa3.5\*\spb5.6-3\*\spa1.7\*\spa3.4\*\spb4.6
-\spa1.3\*\spab7.12.6)}
{4\*\spa1.2\*\spa3.4\*\spb5.6\*\spaba7.12.56.7^3}
\nonumber\\&&
+\frac{\spa1.7\*\spa3.7\*\spab7.12.6\*\spaba1.34.56.7
\*(\spa5.6\*\spb6.4-\spa5.3\*\spb3.4)}
{\spa1.2\*\spaba7.12.56.7^3}\nonumber\\&&
+\frac{\spa1.7^2\*(s_{35}+s_{36}+s_{45}+s_{46})
\*(\spa3.4^2\*\spb4.6^2+\spa3.5^2\*\spb5.6^2)}
{4\*\spa1.2\*\spa3.4\*\spb5.6\*\spaba7.12.56.7^2}
\nonumber\\&&
+\frac{(s_{35}+s_{36}+s_{45}+s_{46})
\*(\spa1.3\*\spab7.12.6+\spa1.7\*\spa3.4\*\spb4.6)^2}
{4\*\spa1.2\*\spa3.4\*\spb5.6\*\spaba7.12.56.7^2}
\nonumber\\&&
-3\*\frac{\spab3.56.4\*\spab5.34.6\*\spaba1.34.56.1
\*(s_{127}-s_{34}+s_{56})
\*(s_{127}+s_{34}-s_{56})}
{4\*\spa1.2\*\spaba7.12.56.7\*\Delta_3(3,4,5,6)^2}
\nonumber\\&&
-5\*\frac{\spa1.7^2\*\spab3.56.4\*\spab5.34.6
\*(s_{127}+s_{34}-s_{56})
\*(s_{127}-s_{34}+s_{56})}
{8\*\spa1.2\*\spaba7.12.56.7^2\*\Delta_3(3,4,5,6)}
\nonumber\\&&
-\frac{\spab3.56.4\*\spab7.12.6\*\spaba5.34.12.7
\*\spaba1.34.56.1}
{\spa1.2\*\spaba7.12.56.7^2\*\Delta_3(3,4,5,6)}
\nonumber\\&&
+\frac{\spa1.3\*\spa5.7\*s_{127}
\*\tspaba1.34.56.7
\*(2\*\spab5.34.6\*\spb4.5
-(s_{345}-s_{346})\*\spb4.6)}
{4\*\spa1.2\*\spaba7.12.56.7^2\*\Delta_3(3,4,5,6)}
\nonumber\\&&
+\frac{\spb3.4\*\tspaba1.34.56.7
\*(2\*\spab3.56.4\*\spa4.5
-(s_{456}-s_{356})\*\spa3.5)
\*(6\*\spa1.3\*\spab7.12.6
+3\*\spa3.4\*\spb4.6\*\spa1.7)}
{4\*\spa1.2\*\spaba7.12.56.7^2\*\Delta_3(3,4,5,6)}
\nonumber\\&&
+\frac{\spa1.3\*(2\*\spab5.34.6\*\spb4.5
-(s_{345}-s_{346})\*\spb4.6)
\*(2\*\spa5.6\*\spab1.34.6-11\*\spa1.5\*s_{127}
+14\*s_{34}\*\spa1.5)}
{8\*\spa1.2\*\spaba7.12.56.7\*\Delta_3(3,4,5,6)}
\nonumber\\&&
+\frac{(2\*\spab3.56.4\*\spa4.5
-(s_{456}-s_{356})\*\spa3.5)
\*(5\*\spb4.6\*\spaba1.34.56.1
-16\*\spa1.3\*\spb3.4\*\spab1.34.6)}
{8\*\spa1.2\*\spaba7.12.56.7\*\Delta_3(3,4,5,6)}
\nonumber\\&&
+(-12\*\spa3.5\*\spb4.6\*\spaba1.34.56.7
+4\*\spa1.5\*\spab7.12.6\*\spab3.27.4\nonumber\\
&-&2\*\spa1.3\*\spb3.4\*\spa3.5\*\spab7.34.6
+3\*\spa1.7\*\spa3.4\*\spb4.6^2\*\spa5.6)
\*\frac{\spa1.7}
{8\*\spa1.2\*\spaba7.12.56.7^2}
\nonumber\\&&
+\frac{\spa1.3\*\spab7.12.6
\*(-4\*\spb1.4\*\spa1.5\*\spa1.7
-2\*\spa1.5\*\spb2.4\*\spa2.7
-9\*\spa1.7\*\spb3.4\*\spa3.5)}
{4\*\spa1.2\*\spaba7.12.56.7^2}
\nonumber\\&&
+\frac{\spa1.3\*\spa1.5\*\spb4.6}
{\spa1.2\*\spaba7.12.56.7}
\Bigg\}
+\Bigg\{\;\Bigg\}_{3 \leftrightarrow 5,4 \leftrightarrow 6}
\end{eqnarray}

\begin{eqnarray}&&
c_{\{34\sep56\}}^{(2)}=\Bigg\{\frac{1}
{4\*s_{34}\*s_{56}}
\*\bigg[\Delta_3(3,4,5,6)
\*\tspbaba6.12.34.56.3\nonumber\\
&\times&\frac{(-\spb2.7\*\spb3.4
\*(\spa3.1\*\spa5.7-\spa3.7\*\spa1.5)
-\spb1.2\*\spa1.5\*\spab1.67.4
+\spb1.2\*\spa1.5\*\spa1.7\*\spb4.7)}
{s_{12}\*\spbab7.56.34.7
\*\spaba7.56.34.7^2}
\nonumber\\&&
-\spa1.6\*\spab5.34.6
\*\tspbaba6.12.34.56.3\nonumber\\
&\times&\frac{(-2\*(s_{356}-s_{456})
\*(\spa1.5\*\spb5.4-\spa1.7\*\spb7.4)
+4\*\spab3.56.4\*(\spa1.5\*\spb5.3-\spa1.7\*\spb7.3))}
{\spa1.2\*\spbab7.56.34.7
\*\spaba7.56.34.7^2}
\nonumber\\&&
-\spa1.5\*(s_{345}-s_{346})
\*\tspbaba6.12.34.56.3\*
\frac{(\spab1.67.4\*(s_{356}-s_{456})
-2\*\spab1.67.3\*\spab3.56.4)}
{\spa1.2\*\spbab7.56.34.7
\*\spaba7.56.34.7^2}
\nonumber\\&&
+\frac{\spa1.5\*\spb2.4\*\spa3.7
\*\spab7.12.6\*\Delta_3(3,4,5,6)}
{s_{12}\*\spaba7.56.34.7^2}
-\frac{\spb1.2\*\spa1.3\*\spa1.5\*\spb4.7\*\spb6.7
\*(s_{345}-s_{346})\*(s_{356}-s_{456})}
{2\*s_{12}\*\spaba7.56.34.7\*\spbab7.56.34.7}
\nonumber\\&&
-\frac{\spa1.3\*\spb4.7\*\Delta_3(3,4,5,6)
\*(
2\*\spab5.12.7\*\spb2.6
+\spb1.2\*\spa1.5\*\spb6.7)}
{2\*s_{12}\*\spaba7.56.34.7\*\spbab7.56.34.7}
-2\*\frac{\spa1.3\*\spab1.56.4\*\spab5.34.6}
{\spa1.2\*\spaba7.56.34.7}
\nonumber\\&&
+\spa1.3\*\spb4.7\*\spab5.34.6
\*(s_{356}-s_{456})
\*\frac{(\spb2.7\*(s_{45}+s_{46}-s_{35}-s_{36})
+2\*\spb1.2\*\spa1.5\*\spb5.7)}
{s_{12}\*\spaba7.56.34.7\*\spbab7.56.34.7}
\nonumber\\&&
+\frac{\spa1.2\*\spb2.7^2\*\spab3.56.4\*\spab5.34.6
\*(s_{35}+s_{45}+s_{36}+s_{46})}
{s_{12}\*\spaba7.56.34.7\*\spbab7.56.34.7}
\nonumber\\&&
-2\*\spa1.3\*\spab3.56.4\*\spab5.34.6
\*\frac{(
\spb1.2\*\spa1.5\*\spb3.7\*\spb5.7
+2\*\spb2.7\*\spb4.7\*\spab4.56.3)}
{s_{12}\*\spaba7.56.34.7\*\spbab7.56.34.7}
\nonumber\\&&
+\frac{\Delta_3(3,4,5,6)
\*\tspbaba6.12.34.56.3
\*\tspbaba4.12.34.56.5}
{\spbab7.56.34.7
\*\spaba7.56.34.7^2}
\*\left(
\frac{2\*\spa1.7^2}
{\spa1.2\*\spaba7.56.34.7}+
\frac{\spb2.7^2}
{\spb1.2\*\spbab7.56.34.7}
\right)
\nonumber\\&&
-\frac{\spaba1.34.56.1
\*(2\*\spab3.56.4\*\spab5.34.6
\*(s_{35}+s_{45}+s_{36}+s_{46})
+\spa3.5\*\spb4.6\*\Delta_3(3,4,5,6))}
{\spa1.2\*\spaba7.56.34.7\*\Delta_3(3,4,5,6)}\bigg]\Bigg\}\nonumber\\
&+&\Bigg\{\;\Bigg\}_{3 \leftrightarrow 5,4 \leftrightarrow 6}
\end{eqnarray}

The result for the helicity flip part of this triangle is,
\begin{equation}
\tilde{c}_{\{34\sep56\}}^{(2)}=\spa3.5\*\spb4.6\*\frac{\spa1.7^2\*\Delta_3(3,4,5,6)-2\*\spaba1.56.34.1\*\spaba7.34.56.7}
{\spa1.2\*s_{34}\*s_{56}\*\spaba7.34.56.7^2}
\end{equation}

\subsubsection{Results for $\{12\sep 56\}$ triangle}
The mass-independent term is obtained by exchange,
\beq
c_{\{12\sep56\}}^{(0)}= \left. c_{\{34\sep56\}}^{(0)} \right|_{1 \leftrightarrow 3, 2 \leftrightarrow 4}
\eeq

The $m^2$ piece contains both symmetric and anti-symmetric parts,
\begin{eqnarray}&&
c_{\{12\sep56\}}^{(2,S)}=
\Bigg\{\frac{1}
{16\*s_{12}\*s_{34}\*s_{56}}\*\bigg[
-4\*\Delta_3(1,2,5,6)
\*\tspbaba6.12.34.56.1\nonumber\\
&\times&\frac{(\spb1.2\*\spb4.7
\*(\spa1.3\*\spa5.7-\spa1.7\*\spa3.5)
+\spb3.4\*\spa3.5
\*(\spab3.67.2+\spa3.7\*\spb7.2))}
{\spbab7.12.56.7\*\spaba7.12.56.7^2}
\nonumber\\&&
-8\*\spb3.4\*\spa3.6\*\spab5.12.6
\*\tspbaba6.12.34.56.1\nonumber\\&&
\*\frac{(s_{156}-s_{256})
\*(\spa3.5\*\spb5.2-\spa3.7\*\spb7.2)
-2\*\spab1.56.2\*(\spa3.5\*\spb5.1-\spa3.7\*\spb7.1)}
{\spbab7.12.56.7\*\spaba7.12.56.7^2}
\nonumber\\&&
-4\*\spb3.4\*\spa3.5\*(s_{125}-s_{126})
\*\tspbaba6.12.34.56.1\*
\frac{(2\*\spab3.67.1\*\spab1.56.2
-\spab3.67.2\*(s_{156}-s_{256}))}
{\spbab7.12.56.7\*\spaba7.12.56.7^2}
\nonumber\\&&
-4\*\frac{\spb3.4\*\spa1.3\*\spb2.6\*\spa3.7\*\spa5.7
\*\Delta_3(1,2,5,6)}
{\spaba7.12.56.7^2}
\nonumber\\&&
-12\*\frac{\spb4.7\*\spab3.12.7\*\spab1.56.2
\*\spab5.12.6\*(s_{15}+s_{16}+s_{25}+s_{26})}
{\spbab7.12.56.7\*\spaba7.12.56.7}
\nonumber\\&&
-4\*\frac{\spa1.3\*\spb2.7\*\spb4.7\*\spab5.12.6
\*(s_{156}-s_{256})
\*(s_{12}+s_{13}+s_{14}+s_{23}+s_{24}
+s_{35}+s_{36}+s_{45}+s_{46})}
{\spbab7.12.56.7\*\spaba7.12.56.7}
\nonumber\\&&
+8\*\frac{\spb4.7\*\spab1.56.2\*\spab5.12.6
\*(-\spab3.12.7\*s_{12}
+\spa1.2\*\spb1.7\*\spb2.7\*\spa3.7
-\spa1.3\*\spb1.7\*(s_{37}+s_{47}))}
{\spbab7.12.56.7\*\spaba7.12.56.7}
\nonumber\\&&
+2\*\frac{\spa1.3\*\spb3.4\*\spb2.7\*\spa3.5\*\spb6.7
\*(\Delta_3(1,2,5,6)-(s_{125}-s_{126})\*(s_{156}-s_{256}))}
{\spbab7.12.56.7\*\spaba7.12.56.7}
\nonumber\\&&
+4\*(\spb2.7\*(s_{156}-s_{256})
\*(\spa3.5\*\spb5.7-\spab3.46.7)
+2\*\spb1.7\*\spab1.56.2\*\spab3.46.7)
\*\frac{\spa1.3\*\spb3.4\*\spab5.12.6}
{\spbab7.12.56.7\*\spaba7.12.56.7}
\nonumber\\&&
+4\*\spb2.7\*\spb4.7\*\spab5.12.6
\*\frac{(2\*\spa1.2\*\spb1.4\*\spa3.4\*\spab1.56.2
-(2\*s_{34}\*\spa1.3+\spa1.2\*\spab3.56.2)\*(s_{156}-s_{256}))}
{\spbab7.12.56.7\*\spaba7.12.56.7}
\nonumber\\&&
-4\*\frac{
\Delta_3(1,2,5,6)
\*\tspbaba6.12.34.56.1
\*\tspbaba2.12.34.56.5}
{\spbab7.12.56.7\*\spaba7.12.56.7^2}
\*\left(\frac{2\*\spa3.7^2\*\spb3.4}
{\spaba7.12.56.7}
+\frac{\spb4.7^2\*\spa3.4}
{\spbab7.12.56.7}\right)
\nonumber\\&&
+4\*\frac{\spb3.4\*\spaba3.56.12.3}
{\spaba7.12.56.7}\*
\left(\frac{
2\*\spab1.56.2\*\spab5.12.6
\*(s_{15}+s_{16}+s_{25}+s_{26})}
{\Delta_3(1,2,5,6)}
+\spa1.5\*\spb2.6\right)
\bigg]
\Bigg\}\nonumber\\
&+&\Bigg\{\;\Bigg\}_{1 \leftrightarrow 5,2 \leftrightarrow 6}
\end{eqnarray}

\begin{eqnarray}
&&c_{\{12\sep56\}}^{(2,A)}=\Bigg\{\frac{1}
{16\*s_{12}\*s_{34}\*s_{56}}
\nonumber\\&&
\*\bigg[
\frac{\spa3.7\*\Delta_3(1,2,5,6)
\*(-2\*\spa1.5\*\spb2.6\*\spab7.12.4
-2\*\spa1.3\*\spb2.6\*\spb3.4\*\spa5.7
-4\*\spb1.2\*\spa1.5\*\spa1.7\*\spb4.6)}
{\spaba7.12.56.7^2}
\nonumber\\&&
+\frac{\spa1.3\*\spb2.4\*\spab5.12.6
\*(-4\*(s_{15}+s_{16}+s_{25}+s_{26})
+8\*(s_{347}-s_{12}))}
{\spaba7.12.56.7}
\nonumber\\&&
-\frac{\spa1.5\*\spb2.7\*\spb6.7
\*(\tspbab7.12.56.4
\*\tspaba7.12.56.3-2\*\spab3.12.4\*\Delta_3(1,2,5,6))}
{\spaba7.12.56.7\*\spbab7.12.56.7}
\nonumber\\&&
+4\*\frac{\spb2.4\*\spa3.5\*\spb6.7
\*\tspaba7.12.56.1}
{\spaba7.12.56.7}\bigg]
\Bigg\}
-\Bigg\{\;\Bigg\}_{1 \leftrightarrow 5,2 \leftrightarrow 6}
\end{eqnarray}

The result for the helicity flip part of this triangle is,
\begin{eqnarray}
&&\label{ct12x56m2}
\tilde{c}_{\{12\sep56\}}^{(2)}=\frac{\spa5.3}
{s_{34}\*s_{56}}
\*\Bigg[-\frac{\spb4.6\*\spa1.7\*\spab7.34.2\*\Delta_3(1,2,5,6)}
{s_{12}\*\spaba7.56.34.7^2}
+\frac{\spb4.6\*\spab1.56.2
\*(s_{15}+s_{16}+s_{25}+s_{26})}
{s_{12}\*\spaba7.56.34.7}\nonumber\\&&
-\frac{\spb4.7\*\tspbaba2.12.34.56.1
\*\tspbab6.12.56.7}
{s_{12}\*\spaba7.56.34.7\*\spbab7.56.34.7}
-2\*\spb2.6\*\frac{(\spb2.4\*(s_{15}+s_{16}+s_{25}+s_{26}+s_{56})
+\spb3.4\*\spab3.56.2)}
{\spb1.2\*\spaba7.56.34.7}\nonumber\\&&
+\frac{\spa1.5\*\spb5.6
\*(2\*\spab1.27.4-4\*\spa1.2\*\spb2.4)}
{\spa1.2\*\spaba7.56.34.7}\Bigg]
\end{eqnarray}

\subsubsection{Results for $\{12\sep 34\}$ triangle}
The results for this triangle are obtained by exchange,
\beq
c_{\{12\sep34\}}^{(i)}= \left. c_{\{12\sep56\}}^{(i)} \right|_{3 \leftrightarrow 5, 4 \leftrightarrow 6} \, ,
\eeq
\beq
\tilde{c}_{\{12\sep34\}}^{(2)}= \left. \tilde{c}_{\{12\sep56\}}^{(2)} \right|_{3 \leftrightarrow 5, 4 \leftrightarrow 6} \, .
\eeq

\subsubsection{Results for $\{7\sep 12\}$ triangle}
The mass-independent part of the coefficient is determined by infrared relations that ensure the cancellation
of $1/\epsilon$ poles in the massless case.  In terms of the box integral coefficients defined above we have,
\beq
   \frac{c_{\{7\sep12\}}^{(0)}}{s_{12}-s_{127}} =
   \frac{d_{\{12\sep56\sep34\}}^{(0)}}{s_{127}\*s_{347}-s_{12}\*s_{34}}
  +\frac{d_{\{12\sep34\sep56\}}^{(0)}}{s_{127}\*s_{567}-s_{12}\*s_{56}} \, .
\eeq
\begin{eqnarray}&&
c_{\{7\sep12\}}^{(2)}=\Bigg\{\frac{1}
{4\*s_{34}\*s_{56}}
\*\Bigg[
4\*\frac{\spa1.7\*\spa3.4\*\spa5.7\*\spb5.6\*\spb2.7
\*\spb4.7^2\*\spab5.12.7}
{\spab7.12.7\*\spaba7.56.34.7
\*\spbab7.56.34.7}
\nonumber\\&&
+4\*\frac{\spa1.7^2\*\spa3.7\*\spa5.7\*\spb4.3
\*(\spb6.5\*\spab5.34.7\*\spa7.3
-\spb6.7\*\spaba7.34.56.3)
\*(s_{35}+s_{36}+s_{45}+s_{46})}
{\spa1.2\*\spaba7.56.34.7^3}
\nonumber\\&&
+\frac{\spa1.7\*\spa5.7\*\spb2.7\*\spb4.7
\*(\spb6.5\*\spab5.34.7\*\spa7.3
-\spb6.7\*\spaba7.34.56.3)
\*(s_{35}+s_{36}+s_{45}+s_{46})}
{\spaba7.56.34.7^2\*\spbab7.56.34.7}
\nonumber\\&&
+2\*\frac{\spa1.7^2\*\spa3.7\*\spab5.34.6\*(\spb4.7
\*(s_{35}+s_{36}+s_{45}+s_{46})
+\spbab4.12.34.7)}
{\spa1.2\*\spaba7.56.34.7^2}
\nonumber\\&&
+\frac{\spb2.6\*\spb2.7\*\spa3.7\*\spa5.7\*\spab7.12.4
\*(s_{35}+s_{36}+s_{45}+s_{46})}
{\spb1.2\*\spaba7.56.34.7^2}
\nonumber\\&&
-\frac{\spa1.7\*\spa3.7\*(2\*\spb2.4\*\spb6.7
\*\spaba5.12.34.7
+\spb1.2\*\spb4.6\*\spaba1.56.12.5)}
{\spaba7.56.34.7^2}
\nonumber\\&&
+\frac{\spa1.7\*\spa3.7\*\spb1.2}
{\spaba7.56.34.7^2}
\nonumber\\&&
\times\Big(\spab1.37.4\*\spab5.34.6
-2\*\spa1.3\*\spb3.4\*\spab5.34.6
+\spa1.2\*\spb2.4\*\spab5.12.6
-\spa1.5\*\spb5.6\*\spab5.36.4\Big)
\nonumber\\&&
+2\*\frac{\spb2.7^2\*\spa5.6\*\spb6.7\*\spab3.12.7\*\spb4.3
\*(\spb6.5\*\spab5.34.7\*\spa7.3
-\spb6.7\*\spaba7.34.56.3)}
{\spb1.2\*\spaba7.56.34.7
\*\spbab7.56.34.7^2}
\nonumber\\&&
+\frac{\spb2.7\*\spb4.7\*\spa3.7\*\spb6.5
\*(-2\*\spab5.34.2\*\spab5.12.7
+3\*\spa1.5\*\spb1.2\*\spab5.12.7
-\spa1.5^2\*\spb1.2\*\spb1.7)}
{\spb1.2\*\spaba7.56.34.7\*\spbab7.56.34.7}
\nonumber\\&&
+\big[\spb2.7\*\spb4.7\*\spa3.7\*(3\*\spa1.5\*\spb6.7
\*(s_{356}+s_{456})
+2\*\spb6.7\*\spaba1.56.12.5\nonumber\\
&+&\spa1.5\*\spa2.5\*\spb2.7\*\spb5.6
-2\*\spa5.7\*\spb6.7\*\spab1.56.7)\big]
\*\frac{1}
{\spaba7.56.34.7\*\spbab7.56.34.7}\nonumber\\&&
-\frac{\spb2.7\*\spa3.7\*\spb4.6
\*(\spa5.3\*\spb3.2+\spa5.4\*\spb4.2-3\*\spa5.6\*\spb6.2)}
{\spb1.2\*\spaba7.56.34.7}\nonumber\\&&
+\frac{\spa1.3\*\spa5.7
\*(\spb2.4\*\spb6.7+3\*\spb2.6\*\spb4.7)}
{\spaba7.56.34.7}\Bigg]\Bigg\} + \Bigg\{\;\Bigg\}_{3 \leftrightarrow 5,4 \leftrightarrow 6}
\end{eqnarray}

\begin{eqnarray}\label{ct712m2}&&
\tilde{c}_{\{7\sep12\}}^{(2)}=\frac{1}
{s_{34}\*s_{56}}
\*\Bigg[\frac{\spa1.7^2\*\spa3.5\*\spb4.6\*\spab7.12.7
\*(s_{35}+s_{36}+s_{45}+s_{46})}
{\spa1.2\*\spaba7.34.56.7^2}
\nonumber\\&&
+\frac{\spab7.12.7}
{\spaba7.34.56.7}
\*\Big(\frac{\spa1.3\*\spa1.5\*\spb4.6}
{\spa1.2}
-\frac{\spb2.4\*\spb2.6\*\spa3.5}
{\spb1.2}\Big)\nonumber\\&&
+\spa3.5\*\spb4.7\*\spb6.7
\*\Big(\frac{\spa1.7^2}
{\spa1.2\*\spaba7.34.56.7}
-\frac{\spb2.7^2}
{\spb1.2\*\spbab7.34.56.7}\Big)\Bigg]
\end{eqnarray}

\subsubsection{Results for $\{7\sep 34\}$ triangle}
The infrared condition here is,
\beq
  \frac{c_{\{7\sep34\}}^{(0)}}{s_{34}-s_{347}}=
  \frac{d_{\{12\sep56\sep34\}}^{(0)}}{s_{127}s_{347}-s_{12}s_{34}}
 +\frac{d_{\{56\sep12\sep34\}}^{(0)}}{s_{347}s_{567}-s_{34}s_{56}} \, .
\eeq

\begin{eqnarray}&&
c_{\{7\sep34\}}^{(2,S)}=\frac{1}
{4\*s_{12}\*s_{34}\*s_{56}}\*
\Bigg\{4\*\frac{s_{34}\*\spa1.2\*\spb5.6
\*\spb2.7^2\*\spa3.7\*\spb4.7\*\spa5.7\*\spab5.34.7}
{\spab7.34.7\*\spaba7.12.56.7
\*\spbab7.12.56.7}
\nonumber\\&&
-4\*\frac{\spa1.7\*\spa3.7^2\*\spa5.7\*\spb1.2\*\spb3.4
\*(\spb6.5\*\spab5.12.7\*\spa7.1
-\spb6.7\*\spaba7.12.56.1)
\*(s_{15}+s_{16}+s_{25}+s_{26})}
{\spaba7.12.56.7^3}
\nonumber\\&&
-2\*\frac{\spb4.7^2\*\spb6.7\*\spab1.34.7
\*\spb1.2\*\spa3.4\*\spa5.6
\*(\spb6.5\*\spab5.12.7\*\spa7.1
-\spb6.7\*\spaba7.12.56.1)}
{\spaba7.12.56.7\*\spbab7.12.56.7^2}
\nonumber\\&&
-\spa3.7\*\spb4.7\*(\spb6.5\*\spab5.12.7\*\spa7.1
-\spb6.7\*\spaba7.12.56.1)
\*(s_{15}+s_{16}+s_{25}+s_{26})
\nonumber\\
&\times&\frac{(2\*\spaba1.34.26.5+\spa1.5
\*(s_{347}+s_{134}+s_{13}+s_{14}))\*\spb1.2}
{\spaba7.12.56.7^2\*\spbab7.12.56.7}
\nonumber\\&&
-\spa3.7\*\spa5.7\*(s_{15}+s_{16}+s_{25}+s_{26})\nonumber\\
&\times&\frac{(\spa1.3\*\spb2.6\*(s_{347}+s_{23}+s_{24})
+\spa1.3\*\spbab2.34.15.6
-2\*\spb2.7\*\spa3.7\*\spab1.25.6)\*\spb4.3}
{\spaba7.12.56.7^2}
\nonumber\\&&
-2\*\frac{\spa3.7^2\*(\spb6.5\*\spab5.12.7\*\spa7.1
-\spb6.7\*\spaba7.12.56.1)
\*(\spb1.2\*\spa1.5-\spb2.6\*\spa5.6)\*\spb4.3}
{\spaba7.12.56.7^2}
\nonumber\\&&
+\frac{\spb2.7\*\spa3.7\*\spb4.7\*\spab5.34.7
\*(-\spa1.5\*\spb5.6\*(s_{15}+s_{16}+s_{25}+s_{26})
+2\*\spa1.2\*\spb2.6\*s_{567})}
{\spaba7.12.56.7\*\spbab7.12.56.7}
\nonumber\\&&
+\frac{\spb2.7\*\spb4.7\*\spab5.34.7
\*(2\*\spb2.3\*\spa3.5\*\spa3.7
+4\*\spb2.4\*\spa3.7\*\spa4.5
-2\*\spb2.4\*\spa3.5\*\spa4.7)\*\spa1.2\*\spb5.6}
{\spaba7.12.56.7\*\spbab7.12.56.7}
\nonumber\\&&
-\frac{\spa1.3\*\spa3.7
\*(\spb2.6\*\spab5.34.7
+2\*\spb2.7\*\spab5.12.6)\*\spb4.3}
{\spaba7.12.56.7}
\nonumber\\&&
+\frac{\spa1.2\*\spb2.6\*\spa3.7
\*(2\*\spb2.4\*\spab5.34.7
+\spb1.2\*\spa1.5\*\spb4.7)}
{\spaba7.12.56.7}\Bigg\}
+\Bigg\{\;\Bigg\}_{1 \leftrightarrow 5,2 \leftrightarrow 6}
\end{eqnarray}

\begin{eqnarray}
&&c_{\{7\sep34\}}^{(2,A)}=\frac{1}
{4\*s_{12}\*s_{34}\*s_{56}}
\*\Bigg\{\frac{\spa3.7\*\spa5.7\*\spab7.34.7\*\spb3.4
\*(\spa1.3\*\spb2.6\*s_{127}+\spb2.4\*\spa3.4\*\spab1.34.6)}
{\spaba7.12.56.7^2}\nonumber\\
&-&\frac{\spa1.5\*\spb2.4\*\spa3.7\*\spab7.34.7
\*(\spab7.34.6\*s_{567}-\spb6.7\*\spaba7.12.56.7)}
{\spaba7.12.56.7^2}\Bigg\}-\Bigg\{\;\Bigg\}_{1 \leftrightarrow 5,2 \leftrightarrow 6}
\end{eqnarray}

The result for the helicity flip part of this triangle is,
\begin{eqnarray}
\tilde{c}_{\{7\sep34\}}^{(2)}&=&
-\spa3.5\*\spb4.6\*\spab7.34.7
\*\frac{(-\spb1.2\*\spa1.7^2\*s_{347}
-2\*\spa1.2\*\spb1.2\*\spa1.7\*\spab7.34.2
+\spa1.2\*\spab7.34.2^2)}
{s_{12}\*s_{34}\*s_{56}\*\spaba7.12.56.7^2}\nonumber\\
&-&\spab7.34.7\*\frac{(-\spb1.2\*\spa1.3\*\spa1.5\*\spb4.6
+\spa1.2\*\spb2.4\*\spb2.6\*\spa3.5)}
{s_{12}\*s_{34}\*s_{56}\*\spaba7.12.56.7}
\end{eqnarray}

\subsubsection{Results for $\{7\sep 56\}$ triangle}
The results for this triangle are obtained by exchange,
\beq
c_{\{7\sep56\}}^{(i)}= \left. c_{\{7\sep34\}}^{(i)} \right|_{3 \leftrightarrow 5, 4 \leftrightarrow 6}
\eeq
\beq
\tilde{c}_{\{7\sep56\}}^{(2)}= \left. \tilde{c}_{\{7\sep34\}}^{(2)} \right|_{3 \leftrightarrow 5, 4 \leftrightarrow 6}
\eeq

\subsection{Results for bubble coefficients}\label{eq:bubble_coefficients}
The bubble coefficients are all independent of the mass. Consequently the bubble integrals do not
contibute to the LR structure. In addition since the full result for the amplitude is UV finite
we have that
\beq
b^{(0)}_{\{12\}}+b^{(0)}_{\{34\}}+b^{(0)}_{\{56\}}+b^{(0)}_{\{127\}}+b^{(0)}_{\{347\}}+b^{(0)}_{\{567\}}=0 \, .
\eeq
\subsubsection{Results for $\{127\}$ bubble}
This function is symmetric under the exchange $(3\leftrightarrow 5,4\leftrightarrow 6)$.
\begin{eqnarray}
&&b^{(0,S)}_{\{127\}}=\Bigg\{
\frac{\spaba1.56.27.1\*\spab5.36.4^2}
{2\*\Delta_3(3,4,5,6)\*\spa1.2
\*\spb3.4\*\spa5.6\*\spaba7.34.56.7}
\nonumber\\&&
-\frac{s_{127}\*\spbab7.34.56.7
\*(\spa1.7\*\spa5.7\*\spab7.12.4)^2}
{\spa1.2\*\spb3.4\*\spa5.6\*\spab7.12.7^2
\*\spaba7.34.56.7^3}
-\frac{\spa1.3\*\spa1.7\*(s_{456}-s_{356})
\*\spab5.34.6\*\spab7.12.4}
{2\*\Delta_3(3,4,5,6)\*\spa1.2\*\spaba7.34.56.7^2}
\nonumber\\&&
+\frac{\spab1.27.4\*\spa1.7\*(s_{127}-s_{56})
\*\spab5.34.6\*\spab7.12.4}
{2\*\Delta_3(3,4,5,6)\*\spa1.2\*\spb3.4
\*\spaba7.34.56.7^2}
-\frac{\spaba1.35.12.7\*\spa1.7
\*\spab3.56.4\*\spab5.34.6}
{2\*\Delta_3(3,4,5,6)\*\spa1.2\*\spaba7.34.56.7^2}
\nonumber\\&&
-\frac{\spa1.5\*\spa1.7\*\spa5.7\*s_{127}
\*(s_{127}-s_{34})\*\spab3.56.4}
{2\*\Delta_3(3,4,5,6)\*\spa1.2\*\spa5.6
\*\spaba7.34.56.7^2}
+\frac{\spa1.5\*\spa1.7\*\spb4.7
\*\spa5.7\*\spab7.12.4}
{2\*\spa1.2\*\spb3.4\*\spa5.6
\*\spaba7.34.56.7^2}
\nonumber\\&&
+\frac{\spa1.7\*\spa5.7\*\spab7.12.4
\*(\spb2.7\*\spb4.6\*\spa5.6
-\spb1.2\*\spa1.5\*\spb4.7
+\spb2.7\*\spb3.4\*\spa3.5)}
{2\*\spb3.4\*\spa5.6\*\spab7.12.7
\*\spaba7.34.56.7^2}
\nonumber\\&&
-\frac{3\*(\spa1.7^2\*\Delta_3(3,4,5,6)
-4\*\spaba1.56.27.1\*\spaba7.34.56.7)
\*\spab3.56.4\*\spab5.34.6\*s_{127}}
{4\*\Delta_3(3,4,5,6)^2
\*\spa1.2\*\spaba7.34.56.7^2}
\nonumber\\&&
+\frac{3\*\spa1.7\*\spb2.7\*\spb4.7
\*\spa5.7^2\*s_{127}\*\spab7.12.4}
{2\*\spb3.4\*\spa5.6\*\spab7.12.7^2
\*\spaba7.34.56.7^2}\Bigg\}
+\Bigg\{\;\Bigg\}_{3 \leftrightarrow 5,4 \leftrightarrow 6}
\end{eqnarray}

\subsubsection{Results for $\{12\}$ bubble}
This function is symmetric under the exchange $(3\leftrightarrow 5,4\leftrightarrow 6)$.
\begin{eqnarray}
b^{(0)}_{\{12\}}&=&\Bigg\{
\frac{\spa5.7^2\*\spab7.12.4^2
\*\tspbaba2.12.34.56.1}
{\spb3.4\*\spa5.6\*\spab7.12.7
\*\spaba7.34.56.7^3}
-\frac{s_{12}\*\spa1.7\*\spb2.7\*\spb4.7\*\spa5.7^2
\*\spab7.12.4}
{2\*\spb3.4\*\spa5.6\*\spab7.12.7^2
\*\spaba7.34.56.7^2}
\nonumber\\&&
-\frac{\spb1.2\*\spa1.7^2\*\spa3.7^2\*\spa5.7^2
\*\spbab7.34.56.7}
{2\*\spa3.4\*\spa5.6\*\spab7.12.7
\*\spaba7.34.56.7^3}
-\frac{\spa1.2\*\spb2.7^2\*\spa3.7^2\*\spa5.7^2}
{2\*\spa3.4\*\spa5.6\*\spab7.12.7
\*\spaba7.34.56.7^2}
\nonumber\\&&
-\spa1.7\*\spa5.7\*\spb1.2\*\spb4.7
\*\frac{(-3\*\spa1.5\*\spab7.12.4+\spa1.7\*\spab5.12.4)}
{2\*\spb3.4\*\spa5.6\*\spab7.12.7
\*\spaba7.34.56.7^2}
+\frac{\spa1.2\*\spa5.7^2\*\spb2.4^2}
{\spb3.4\*\spa5.6\*\spaba7.34.56.7^2}
\nonumber\\&&
+\frac{\spa1.2\*\spb2.4^2\*\spaba5.34.12.5}
{2\*\spb3.4\*\spa5.6\*\spaba7.34.56.7
\*\Delta_3(1,2,3,4)}
+\frac{\spa1.2\*\spb2.4\*\spa5.7\*\spab5.12.4
\*\spab7.34.2\*(s_{12}-s_{567})}
{2\*\spb3.4\*\spa5.6\*\spaba7.34.56.7^2
\*\Delta_3(1,2,3,4)}
\nonumber\\&&
+\frac{s_{12}\*\spa3.5^2\*\spab1.34.2}
{2\*\spa3.4\*\spa5.6\*\spaba7.34.56.7
\*\Delta_3(1,2,3,4)}
+\frac{s_{12}\*(s_{12}-s_{567})\*\spa1.7
\*\spa3.5\*\spa3.7\*\spab5.34.2}
{2\*\spa3.4\*\spa5.6\*\spaba7.34.56.7^2
\*\Delta_3(1,2,3,4)}
\nonumber\\&&
+\frac{\spa1.5\*\spa3.7\*(\spb1.2\*\spa1.5\*\spa3.7
-2\*\spb1.2\*\spa1.7\*\spa3.5
+2\*\spb2.4\*\spa3.4\*\spa5.7)}
{2\*\spa3.4\*\spa5.6\*\spaba7.34.56.7^2}
\nonumber\\&&
-\frac{3\*\spab1.34.2\*\spab3.12.4
\*\spaba5.34.12.5\*(s_{34}-s_{567}-s_{12})}
{\spa5.6\*\spaba7.34.56.7\*\Delta_3^2(1,2,3,4)}
\nonumber\\&&
+\frac{\spa1.7\*\spaba5.34.12.5
\*(\spa3.7\*\spb1.2\*\spab1.23.4
-\spa1.3\*\spab7.12.4\*\spb1.2
+7\*\spab3.12.4\*\spab7.56.2)}
{2\*\spa5.6\*\spaba7.34.56.7^2\*\Delta_3(1,2,3,4)}
\nonumber\\&&
+\big\{
-2\*\spb1.2\*\spa1.7\*\spb3.4\*\spa3.5
-4\*\spa1.5\*\spb2.4\*\spab7.34.1
-4\*\spb2.4\*\spa2.5\*\spab7.34.2
\nonumber\\&&
+\spa3.5\*\spb3.4\*\spab7.34.2
-2\*\spa3.4\*\spa5.7\*\spb2.4\*\spb3.4\big\}
\nonumber\\&&
\times\frac{\spb1.2\*\spa1.2\*\spa1.7\*\spa3.5}
{2\*\spa5.6\*\spaba7.34.56.7^2\*\Delta_3(1,2,3,4)}
\nonumber\\&&
+\spa3.7\*\tspaba1.12.34.7
\*\frac{(-\spb1.2\*\spa1.5\*\spab5.12.4
-2\*\spb2.4\*\spaba5.34.67.5)}
{2\*\spa5.6\*\spaba7.34.56.7^2\*\Delta_3(1,2,3,4)}
\nonumber\\&&
+\Big(2\*\spa1.3\*\spa1.7\*\spb1.2\*\spab5.12.4
+2\*\spa1.5\*\spab3.12.4\*\spab7.56.2
+\spa1.2\*\spa3.5\*\spb2.4\*\spab7.56.2\Big)
\nonumber\\&&
\times\frac{\tspaba5.12.34.7}
{2\*\spa5.6\*\spaba7.34.56.7^2\*\Delta_3(1,2,3,4)}
\nonumber\\&&
+\Big\{-7\*\spb2.4\*\spa2.5\*\spa3.5\*\spab1.34.2
+\spb2.4^2\*\spa1.2\*\spa4.5\*\spa3.5
+\spa3.5^2\*\spb2.4\*\spa1.2\*\spb2.3
\nonumber\\&&
+\spa1.5\*\spa2.3\*\spb2.4\spb1.2\*\spa1.5
+\spb1.2\*\spa1.3\*\spa1.5
\*(\spb1.4\*\spa1.5-2\*\spb3.4\*\spa3.5)
\nonumber\\&&
-7\*\spb2.4\*\spa3.5\*\spa1.5\*(s_{13}+s_{14})
+5\*\spa1.5\*\spb2.4\*\spa3.4\*\spab5.12.4\Big\}
\nonumber\\&&
\times\frac{1}
{2\*\spa5.6\*\spaba7.34.56.7\*\Delta_3(1,2,3,4)}
\Bigg\}
\nonumber\\&&
+\Bigg\{\;\Bigg\}_{3\leftrightarrow5,4\leftrightarrow6}
\end{eqnarray}

\subsubsection{Results for remaining bubble coefficients}
The remaining bubble coefficients are all obtained by exchange,
\begin{eqnarray}
b^{(0)}_{\{347\}}&=& \left. b^{(0)}_{\{127\}} \right|_{1 \leftrightarrow 3, 2 \leftrightarrow 4},\;\;
b^{(0)}_{\{567\}}= \left. b^{(0)}_{\{127\}} \right|_{1 \leftrightarrow 5, 2 \leftrightarrow 6}\, , \\
b_{\{34\}}^{(0)}&=& \left. b_{\{12\}}^{(0)} \right|_{1 \leftrightarrow 3, 2 \leftrightarrow 4}\,,\;\;
b_{\{56\}}^{(0)}= \left. b_{\{12\}}^{(0)} \right|_{1 \leftrightarrow 5, 2 \leftrightarrow 6}\,.
\end{eqnarray}

\subsubsection{Results for rational term}
The rational piece is determined by the triangle coefficients proportional to $m^2$
and the unique box coefficient proportional to $m^4$, (see Eqs.~(\ref{d4one},\ref{d4two},\ref{d4three})) 
\beq
r=\frac{1}{2}(c_{\{7\sep12\}}^{(2)}+c_{\{7\sep34\}}^{(2)}+c_{\{7\sep56\}}^{(2)}
      +c_{\{34\sep56\}}^{(2)}+c_{\{12\sep56\}}^{(2)}+c_{\{12\sep34\}}^{(2)}
      -d_{\{12\sep34\sep56\}}^{(4)}) \, .
\eeq

\section{Numerical implementation}
\label{timing}

We can directly compare our analytic results against those obtained
using Recola2~\cite{Denner:2017wsf},
with the model file `{\tt SM\_FERM}' that only computes the effects of
fermion loops.  In order to perform a direct comparison we must also
account for additional contributions to the amplitude from triangle diagrams
in which the $Z$ bosons are not both attached to the quark loop.
Representative diagrams for these three processes are shown in
Fig.~\ref{trianglediags} and explicit results for the contributions are given
in Appendix~\ref{anomtriangles}.

We have performed a comparison for the $ZZ$+jet process, 
obtaining perfect agreement. For the purposes of this comparison
we have used a physical value of the top-quark mass and set the bottom
quark mass to zero.  We observe
that the analytic amplitudes presented here are more than an order of magnitude
faster than their Recola2 counterparts.

\begin{figure}
\includegraphics[width=0.375\textwidth,angle=270]{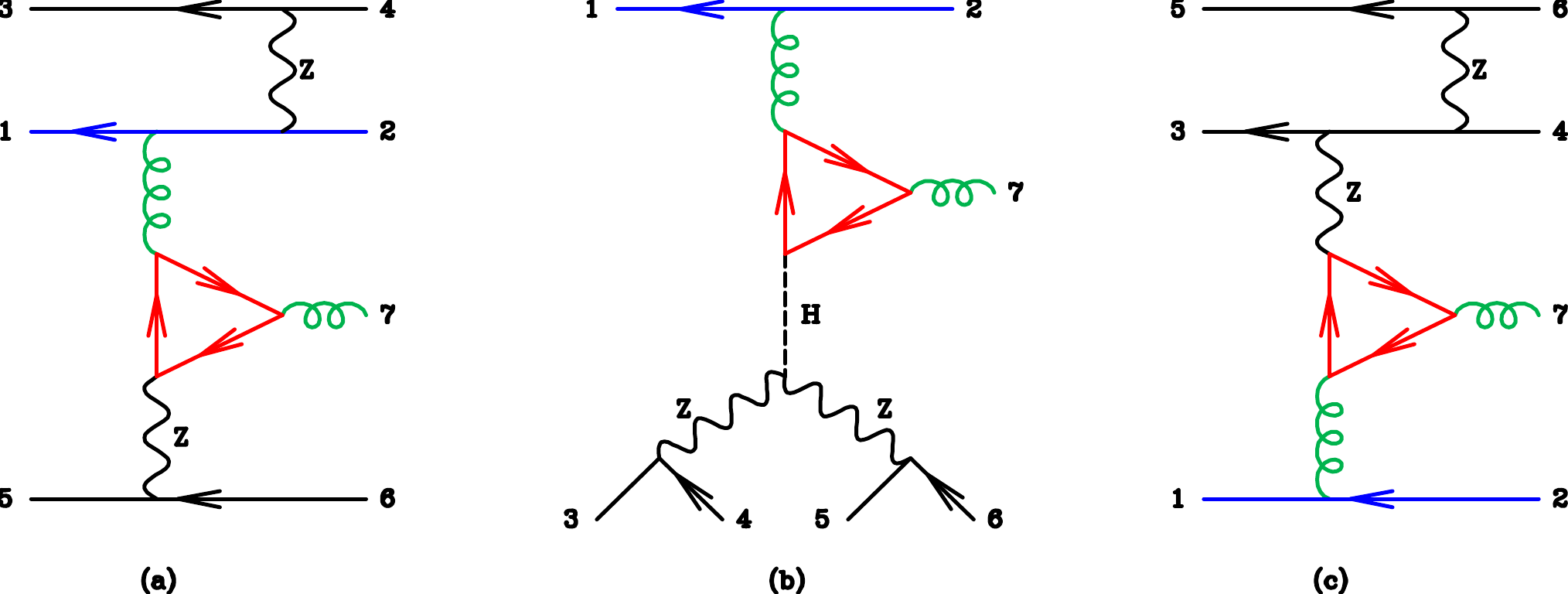}
\caption{Representative diagrams for additional processes
  (a) Doubly resonant anomaly diagram;
  (b) Higgs boson contribution;
  (c) Singly resonant anomaly diagram.
  \label{trianglediags}}
\end{figure}

\section{Conclusions}
\label{conclusions}

In this paper, we have presented compact analytical expressions for
the rational coefficients of the master integrals of the one-loop QCD
helicity amplitudes for the production of a pair of vector bosons in
association with a jet. We have focused on the contribution mediated
by a closed quark loop, and retained full dependence on the quark mass. The
results are expressed in spinor-helicity variables by factoring out
propagators involving masses and by expressing the results in terms of
the massless decay products of the vector bosons, which are therefore
considered fully off mass shell.

Due to the large number of scales involved in 7-point phase space and
the fact that the rational coefficients are ratios of
polynomials subject to constraints, namely momentum-conservation and
Schouten identities, simplifying the analytical expressions is a
complex task. To tackle it, we rely and expand upon recent advances in
spinor-helicity simplification techniques based on algebraic geometry
and numerical sampling in singular limits. In particular, for the
first time beyond five-point amplitudes, we systematically identify
irreducible varieties in spinor space, thus quantitatively identifying
the pole structure of the amplitude in the analytical continuation to
complex momenta. We also observe that some of the involved ideals are
not radical, meaning special care is required in making the connection
between numerical evaluations and membership to symbolic powers. We
employ floating-point and $p$-adic evaluations close to these
irreducible varieties to infer membership of numerator polynomials to
symbolic powers of prime ideals. Subsequently, this allows to identify
possible partial-fraction decompositions as well as new numerator
structures. We fit the numerators by sampling near singular varieties.

The usefulness of compact analytical expressions, and related
simplification techniques, goes beyond that of mere theoretical
understanding. As the computational load on the Worldwide LHC
Computing Grid is predicted to fall short of the demands in the near
future~\cite{HEPSoftwareFoundation:2017ggl},
speeding up matrix-element providers by using simplified analytic
expressions~\cite{Campbell:2021vlt} would improve the
performance of event generators and aid LHC data
analysis. Furthermore, in light of phenomenological applications, and
in particular for the numerical stability in singular regions, it
would be interesting to understand the interplay between the primary
decompositions in complexified momentum space and the real subset of
the latter needed for physical kinematics.

\acknowledgments
We thank Ben Page for useful discussions and comments on the draft.
This manuscript has been authored by Fermi Research Alliance, LLC
under Contract No. DE-AC02-07CH11359 with the U.S. Department of
Energy, Office of Science, Office of High Energy Physics.

\appendix
  
\section{Triangle contributions}
\label{anomtriangles}

This section presents results for the triangle contributions illustrated
in Fig.~\ref{trianglediags}.
There are three types of triangle contributions to enumerate:
\begin{enumerate}
\item Axial anomaly diagrams with one $Z$ boson coupling to the external quark line
and the other to the fermion loop, Fig.~\ref{trianglediags}(a).
\item Higgs-mediated contributions with the Higgs boson coupling to the
fermion loop, Fig.~\ref{trianglediags}(b).
\item Single-resonant axial anomaly diagrams, with one $Z$ boson coupled to the fermion
loop whose decays products subsequently radiate the second $Z$ boson, Fig.~\ref{trianglediags}(c).
\end{enumerate}
These contributions all
take a very simple form since they can be obtained by contracting suitable
currents with known results for triangle loops containing two off-shell
gluons and either a $Z$ or a Higgs boson.

\subsection{Double resonant axial anomaly}

We first consider contributions such as those depicted in Fig.~\ref{trianglediags}(a).
The amplitude for the production of a $Z$ boson by two offshell gluons has been given
for example in ref.~\cite{Campbell:2016tcu} and, more specifically for the case at hand,
in Appendix A of ref.~\cite{Campbell:2007ev}.  The amplitude for this
contribution is given by contracting this result with the appropriate currents. 
The basic amplitude is,
\begin{eqnarray}
A_{ax,56}(1^-,2^+,3^-,4^+,5^-,6^+,7^+) &=& \frac{\mathcal{G} \spb7.6}{\stw \ctw s_{34} s_{56}} \nn \\
 &\times & \left(\frac{\spab5.13.4 \spb2.7 \spa1.3}{s_{134}}
      -\frac{\spa5.1 \spab3.24.7 \spb4.2}{s_{234}}\right) \, ,
\end{eqnarray}
with other helicity amplitudes obtained trivially by symmetries. For example,
\beq
A_{ax,56}(1^-,2^+,3^-,4^+,5^-,6^+,7^-) = \left. - A_{ax,56}(2^-,1^+,4^-,3^+,6^-,5^+,7^+)
 \right|_{\spa.. \leftrightarrow \spb..} \,,
\eeq
\beq
A_{ax,56}(1^-,2^+,3^-,4^+,5^+,6^-,7^+) = A_{ax,56}(1^-,2^+,3^-,4^+,6^-,5^+,7^+) \,,
\eeq
\beq
A_{ax,56}(1^-,2^+,3^+,4^-,5^-,6^+,7^+) = A_{ax,56}(1^-,2^+,4^-,3^+,5^-,6^+,7^+) \,,
\eeq
\beq
A_{ax,56}(1^+,2^-,3^+,4^-,5^-,6^+,7^+) = -A_{ax,56}(2^-,1^+,3^-,4^+,5^-,6^+,7^+) \,.
\eeq
The loop integral factor $\mathcal{G}$ is defined by
$\mathcal{G} = F_1(p_{1234},p_{56};m_t) - F_1(p_{1234},p_{56};m_b)$ where,
\beq
F_1(p_1, p_2; m) = \frac{1}{2(p_2^2-p_1^2)} \left[
 1+ 2m^2 C_0(p_1, p_2;m)
 + \frac{p_2^2}{p_2^2-p_1^2}
 (B_0(p_2^2;m) - B_0(p_1^2;m)) \right]
\label{eq:F1defn}
\eeq
Including the overall factors and also accounting for the additional contribution where the
vector boson decay products ($(3,4)$ and $(5,6)$) are
interchanged we have,
\begin{eqnarray}
{\cal A}_7^{ax, B} = 4 i e^4 \frac{g_s}{16 \pi^2}  t^B &\Big[&
   P_Z(e_q,q_{34},v_q,v_{34}) P(s_{56},M_Z) v_{56} A_{ax,56}(1,2,3,4,5,6,7) \nn \\
&+&P_Z(e_q,q_{56},v_q,v_{56}) P(s_{34},M_Z) v_{34} A_{ax,56}(1,2,5,6,3,4,7) \Big]
\end{eqnarray}

\subsection{Higgs contribution}

We now address another component of the vector boson 
pair production amplitude, which is the piece containing an intermediate Higgs boson as shown
in Fig.~\ref{trianglediags}(b). This result has been known for almost 35 years~\cite{Ellis:1987xu,Baur:1989cm}.
The relevant amplitude is,
\beq
0 \to q(p_1) +\qbar(p_2)+H(VV)+g(p_7)\, .
\eeq
This process is of interest since it is one of the simplest processes
to illustrate the fundamental role of the Higgs boson in cancelling
bad high energy behaviour~\cite{Lee:1977eg}.
The amplitude for the production of a
Higgs boson by two offshell gluons has been given for example in
refs.~\cite{Campbell:2014gua,Budge:2020oyl}.
Using that result and attaching the decays of the $Z$ bosons we obtain,
\begin{eqnarray}
\label{eq:Hamp}
&&{\cal A}_7^{h,B}(1^-,2^+,3^-,4^+,5^-,6^+,7^+) = 4 i e^4 \frac{g_s^2}{16 \pi^2} t^B \\  &&
   \frac{F_T \, l_{34} l_{56}}{4 \sstw \cstw} \frac{P(s_{127},M_H)}{s_{127}}
    \frac{\spa1.2 \spb2.7^2}{(s_{127}-s_{12})}
    \frac{\spa3.5 \spb6.4}{s_{12}}
    \frac{P(s_{34},M_Z)}{s_{34}}
    \frac{P(s_{56},M_Z)}{s_{56}} \nn \, ,
\end{eqnarray}
where
\begin{eqnarray}
  F_T &=&\frac{4m^2}{s_{127}-s_{12}} \Bigg[1+\frac{s_{12}}{s_{127}-s_{12}} \big(B_0(s_{127};m)- B_0(s_{12};m)\big) \nonumber \\
    &+&\left(2 m^2-\frac{(s_{127}-s_{12})}{2} \right) C_0(p_{12},p_7;m)\Bigg]\, ,
\end{eqnarray}
and the coupling factors $l_{34}$ and $l_{56}$ are the left-handed couplings of the $Z$ boson
decay products (equal to either $v_{L,e}$ or $v_{L,n}$ in
Eqs.~\eqref{leptoncouplings} and~\eqref{leptoncouplings2}).
Amplitudes for quarks and leptons of opposite helicity are trivially obtained by interchanging
labels ($1 \leftrightarrow 2$, $3 \leftrightarrow 4$ or $5 \leftrightarrow 6$) and coupling factors.
The amplitude for a negative helicity gluon is obtained by making the replacement
$\spa1.2 \spb2.7^2 \to \spb1.2 \spa1.7^2$ in Eq.~\eqref{eq:Hamp}.

\subsection{Single resonant axial anomaly}

This contribution corresponds to diagrams such as the one in Fig.~\ref{trianglediags}(c).
As in the double-resonant case, the result for this contribution can be obtained by contracting
the appropriate currents with the known result from refs.~\cite{Campbell:2007ev,Campbell:2016tcu}.
In this case there are two essential amplitudes,
\begin{eqnarray}
&& A_{sr}(1^-,2^+,3^-,4^+,5^-,6^+,7^+) =
 \frac{\mathcal{F}}{\stw\ctw} \frac{P(s_{127},M_Z)}{s_{127}} \spb2.7 \Big[ \nn \\ &&
   l_{34} \frac{P_Z(q_{34},q_{56},l_{34},l_{56},s_{56})}{s_{56}}   \Big( 
       \frac{\spa1.3 \spb4.6 \spab5.4+6.7}{s_{456}}  
     + \frac{\spa3.5 \spb4.7 \spab1.3+5.6}{s_{356}} \Big) \nn \\ &&
 - l_{56} \frac{P_Z(q_{34},q_{56},l_{34},l_{56},s_{34})}{s_{34}}   \Big(
       \frac{\spa1.5 \spb4.6 \spab3.6+4.7}{s_{346}} 
     + \frac{\spa3.5 \spb6.7 \spab1.5+3.4}{s_{345}} \Big) \Big] \, ,
\end{eqnarray}
and,
\begin{eqnarray}
&& A_{sr}(1^-,2^+,3^-,4^+,5^+,6^-,7^+) =
 \frac{\mathcal{F}}{\stw\ctw} \frac{P(s_{127},M_Z)}{s_{127}} \spb2.7 \Big[ \nn \\ &&
   l_{34} \frac{P_Z(q_{34},q_{56},l_{34},r_{56},s_{56})}{s_{56}}   \Big( 
    \frac{\spa1.3 \spb4.5 \spab6.4+5.7}{s_{456}}
  + \frac{\spa3.6 \spb4.7 \spab1.36.5}{s_{356}} \Big) \nn \\ &&
 + r_{56} \frac{P_Z(q_{34},q_{56},l_{34},r_{56},s_{34})}{s_{34}}   \Big(
    \frac{\spa1.6 \spb4.5 \spab3.45.7}{s_{345}}
  + \frac{\spa3.6 \spb5.7 \spab1.36.4}{s_{346}} \Big) \Big]\, .
\end{eqnarray}
We note that, even accounting for coupling changes,
the two are not related by a $5 \leftrightarrow 6$
interchange because one of the terms flips sign.
These amplitudes depend explicitly on the charges of the decay products
($q_{34}$, $q_{56}$) and their left- ($l_{34}$, $l_{56}$)
and right-handed ($r_{34}$, $r_{56}$) couplings to $Z$ bosons. The
combination of couplings and propagator factors is,
\beq
P_Z(q_1,q_2,v_1,v_2,s) = q_1q_2 + v_1 v_2 \, P(s,M_Z) \,.
\eeq
The loop factor is
$\mathcal{F} = F_1(p_{12},p_{3456},m_t) - F_1(p_{12},p_{3456},m_b)$
where $F_1$ has already been specified in Eq.~\eqref{eq:F1defn}.
Remaining amplitudes are obtained by symmetry operations.  These correspond to flipping
the helicities of the quarks, e.g.,
\beq 
A_{sr}(1^+,2^-,3^-,4^+,5^+,6^-,7^+) =  A_{sr}(2^-,1^+,3^-,4^+,5^+,6^-,7^+) \,,
\eeq
flipping the helicities of the leptons, e.g.
\beq 
A_{sr}(1^-,2^+,3^+,4^-,5^-,6^+,7^+) = \left.  - A_{sr}(1^-,2^+,4^-,3^+,6^+,5^-,7^+)
 \right|_{l_{34} \to r_{34}, r_{56} \to l_{56}} \,,
\eeq
and flipping the gluon helicity, e.g.
\beq 
A_{sr}(1^-,2^+,3^-,4^+,5^+,6^-,7^+) = \left. - A_{sr}(2^-,1^+,4^-,3^+,6^+,5^-,7^+)
 \right|_{\spa.. \leftrightarrow \spb..} \,.
\eeq
Including overall factors we then have,
\begin{eqnarray}
{\cal A}_7^{sr,B} = 4 i e^4 \frac{g_s^3}{16 \pi^2}  t^B A_{sr}\, .
\end{eqnarray}

\section{Integrals}
\label{Integrals}
This section gives the precise definition of the scalar integrals.
We define the denominators of the integrals as follows,
\beq \label{denominatordef}
D(\ell) = \ell^2-m^2+i\varepsilon \, .
\eeq
Either the denominators all have a common non-zero mass, for the case of $ZZ$ production,
or the mass can be taken to be zero, $m=0$ for the case of $WW$ production. In the latter
case we ignore the contributions of top and bottom loops. 
The momenta running through the propagators are,
\begin{eqnarray} \label{denominators}
\ell_1 &=& \ell+p_1 = \ell +q_1 \, \nonumber \\
\ell_{12} &=& \ell+p_1+p_2 = \ell +q_2 \, \nonumber \\
\ell_{123}&=& \ell+p_1+p_2+p_3 = \ell +q_3 \, \nonumber \\
\ell_{1234} &=& \ell+p_1+p_2+p_3+p_4 = \ell +q_4 \, .
\end{eqnarray}
The $p_i$ are the external momenta, whereas the $q_i$ are the off-set momenta in the propagators.
In terms of these denominators the integrals are,
\begin{eqnarray}
\label{Integral_defns}
B_0(p_1;m) &=& \frac{\bar\mu^{4-n}}{r_\Gamma}\frac{1}{i \pi^{n/2}} \int {\rm d}^n\ell \,\frac{1}{D(\ell)\,D(\ell_1)}\, , \nonumber \\
C_0(p_1,p_2;m) &=& \frac{1}{i \pi^{2}} \int {\rm d}^4\ell \,\frac{1}{D(\ell)\,D(\ell_1)\,D(\ell_{12})}\, ,\nonumber \\
D_0(p_1,p_2,p_3;m) &=& \frac{1}{i \pi^{2}} \int {\rm d}^4\ell \,\frac{1}{D(\ell)\,D(\ell_1)\,D(\ell_{12})\,D(\ell_{123})}\, .
\end{eqnarray}
where $r_\Gamma=1/\Gamma(1-\epsilon)+O(\epsilon^3)$ and $\bar\mu$ is an arbitrary mass scale.

\section{Asymmetric approaches and ring extensions}\label{sec:AsymmetricApproachesAndRingExtensions}

In Section~\ref{sec:primary-decompositions} we presented several
primary decompositions, some of which involve ideals which are not
radical, while
Eq.~\eqref{eq:maximal-codimension-symbolic-power-membership}, which is
the basis for the partial-fraction decomposition of
Eq.~\eqref{eq:partial-fraction-decomposition}, requires the involved
ideal $\left\langle \mathcal{D}_\alpha, \mathcal{D}_\beta \right\rangle$
to be radical. A natural question which arises is then
how to generalize this to non-radical ideals. Furthermore, even when
the condition on Eq.~\eqref{eq:maximal-codimension-symbolic-power-membership}
is satisfied, i.e.~when the ideal is radical, the phase-space point
chosen in Eq.~\eqref{eq:symmetric-approach-point} is a very specific
one, which satisfies $\mathcal{D}_\alpha \sim \mathcal{D}_\beta \sim \epsilon$.
In this case $\mathcal{D}_\alpha, \mathcal{D}_\beta$ vanish symmetrically.
It was shown in ref.~\cite{DeLaurentis:2019phz} that by allowing unequal
(i.e.~asymmetric) degrees of vanishing it is possible glean more
information on the numerators. In this appendix, we address both
points in unified way with algebraic geometry. In particular, we are
going to reduce the asymmetric case to the symmetric one and, in doing
so, provide an algebraic interpretation to evaluations in asymmetric
approaches.

First of all, let us generalize
Eq.~\eqref{eq:maximal-codimension-symbolic-power-membership} to a
special class of non-radical ideals
$\left\langle \mathcal{D}_\alpha, \mathcal{D}_\beta \right\rangle$,
specifically to those whose primary
components are (symbolic) powers of their associated
primes\footnote{By definition of primary, if a primary ideal is a
  power of the associated prime ideal, then this power is also a
  symbolic power.}.
To achieve this, let us recall that the
$\kappa^{\text{th}}$ symbolic power $Q^{\langle \kappa \rangle}$ of any
$P$-primary ideal $Q$ can be defined as the $P$-primary component of $Q^{\kappa}$,
that is,
\beq
Q^{\kappa} = Q^{\langle \kappa \rangle} \cap Q^{\text{em.}}_1 \cap \dots \cap Q^{\text{em.}}_m \, ,
\eeq
where $Q^{\langle \kappa \rangle}$ is $P$-primary and the
$Q^{\text{em.}}_{1\leq i \leq m}$ are embedded. See, for instance,
ref.~\cite[Lemma 1.18]{grifo_2018} for why this is a valid definition. It follows that,
\beq
Q^{\langle \kappa \rangle} = P^{\langle s \kappa \rangle} \quad \text{if} \quad Q = P^{s} \, .
\eeq
We denote the exponent as $s$ because it is the saturation index of
$P$ in $Q$. Now, let the primary decomposition of $\left\langle
\mathcal{D}_\alpha, \mathcal{D}_\beta \right\rangle$ read,
\beq
\left\langle \mathcal{D}_\alpha, \mathcal{D}_\beta \right\rangle = \bigcap_l Q_l = \bigcap_l P_l^{s_l} \, ,
\eeq
and define $\kappa$ such that,
\beq
\label{eq:generalized-kappa-definition}
  \kappa = \text{min}\big( \kappa_l \, : \, \mathcal{N}_i \; \text{vanishes to order} \; s_l \cdot \kappa_l \; \text{on} \; V(Q_l) \big) \, .
\eeq  
We have then achieved a generalization of
Eq.~\eqref{eq:maximal-codimension-symbolic-power-membership}, it now
reads,   
\beq
\label{eq:generalized-maximal-codimension-symbolic-power-membership}
  \mathcal{N}_i \in \bigcap_l P_l^{\langle s_l \cdot\kappa\rangle}= \bigcap_l Q_l^{\langle\kappa\rangle}\;\, \Rightarrow \;\, \mathcal{N}_i \in \big\langle \mathcal{D}_\alpha , \mathcal{D}_\beta \big\rangle ^{\langle \kappa \rangle} \, , \;\, \text{if} \;\, Q_l = P_l^{s_l} \, .
\eeq
This generalization is, however, insufficient: not a single
non-radical primary ideal obtained in Section~\ref{sec:primary-decompositions}
is a power of the associated prime.
Since the prime ideal $P_l = \displaystyle\sqrt{Q_l}$ is unique, if no
positive integer $s_l$ exists such that $Q_l = P_l^{s_l}$, then it is
not possible to find the desired prime ideal $P_l$ in $R_n$ such that
the condition on
Eq.~\eqref{eq:generalized-maximal-codimension-symbolic-power-membership}
is satisfied. Nevertheless, even if such a $P_l$ does not exist in
$R_n$, we can achieve this by extending the ring in which the ideal is
defined by allowing it to include roots of polynomials.

For our purposes, it suffices to consider a ring extension involving a
single $s^{\text{th}}$-root. This can be achieved by extending the quotient
ring $R_n$ of Eq.~\eqref{eq:quotient_ring} by a single variable $x$,
and by taking the quotient with respect to an ideal defining $x^s$ as
a member $q$ of $R_n$. That is, we define the extended ring
$R_n^\sharp$ as,
\beq
R_n^\sharp = R_n\kern-0.8mm\left[x\right] \big/ \left\langle x^s - q \right\rangle_{R_n\left[x\right]}  \, ,
\eeq
where $R_n\kern-0.8mm\left[x\right]$ denotes the ring of polynomials
in $x$ with coefficients in $R_n$.  Given this definition,
$R_n^\sharp$ is a quotient ring of (an extension of) a quotient
ring. Nevertheless, we can also regard $R_n^\sharp$ as a simple
quotient ring, just like $R_n$. In fact, by the
\textit{Third Isomorphism Theorem} \cite{cox2006using}, we have,
\beq
R_n^\sharp \cong S_n\kern-0.8mm\left[ x \right] \big/ \left\langle \sum_{i=1}^n |i⟩[i|, x^s - q \right\rangle_{S_n\left[ x \right]} \, .
\eeq    
That is, $R_n^\sharp$ is isomorphic to the quotient of the (extended) polynomial
ring $S_n\kern-0.8mm\left[ x \right]$, by an ideal whose generators
define both momentum conservation and $x^s$ as a polynomial in
$S_n$. Since this ideal is of maximal codimension, by the same
reasoning which makes $R_n$ a Cohen--Macaulay ring
\cite{DeLaurentis:2022otd}, $R_n^\sharp$ is also Cohen--Macaulay.
Effectively, we have extended $R_n$ by $\sqrt[s]{q}$. Therefore, with a slight
abuse of notation, let us simply denote $x$ as $\sqrt[s]{q}$. In the
following, $R_n^\sharp$ will denote different extensions of $R_n$, but
it will always be clear which one is being considered at any one time
depending on the polynomial appearing under the root. We are now in a position
to build an ideal $P^\sharp$ of $R^\sharp_n$ such that,
\beq\label{eq:primary-as-power-of-prime-in-extension}
(P^\sharp)^{s} \cap R_n = Q \, ,
\eeq
even if no prime ideal $P$ of $R_n$ exists such that $P^{s}=Q$. Note
that using the symbolic power instead of the standard power in
Eq.~\eqref{eq:primary-as-power-of-prime-in-extension} would again be
redundant: either the symbolic power coincides with the standard
power, or the embedded components in the primary decomposition of the
standard power must become redundant in the intersection with
$R_n$\footnote{To show this, one has to remember that $Q$ is a primary
  ideal, and that inclusion is preserverd in the intersection with a
  subring, i.e.~$\sqrt{(P^\sharp)^{\langle s \rangle}} \subset
  \sqrt{Q^{\sharp, \text{em.}}_i} \, \Rightarrow \, \sqrt{(P^\sharp)^{\langle s \rangle}} \cap R_n \subseteq
  \sqrt{Q^{\sharp, \text{em.}}_i} \cap R_n$.}.  By the same reasoning, given
Eq.~\eqref{eq:primary-as-power-of-prime-in-extension}, it can be shown that,
\beq
(P^\sharp)^{\langle s \kappa \rangle} \cap R_n = Q^{\langle \kappa \rangle} \, .
\eeq
As the numerators $\mathcal{N}_i$ belong to the quotient ring $R_n$,
we are always free to add the intersection with $R_n$ to a membership
statement of the form $\mathcal{N}_i \in P^{\sharp \, \langle s \kappa
  \rangle}$. Thus, assuming we can find an appropriate $P^\sharp$ and
up to a suitable re-scaling of the symbolic power, we are now in a
position to numerically obtain information about membership to
$Q^{\langle \kappa \rangle}$, independently of whether $Q$ is the
power of a prime ideal in $R_n$. That is, we have completely
generalized
Eq.~\eqref{eq:generalized-maximal-codimension-symbolic-power-membership}. Furthermore,
by extending the reasoning to $R_n^\sharp$, we have provided an
interpretation to the $\kappa^{\text{th}}$ symbolic power of a
non-radical ideal $I = \bigcap_l Q_l$ free of embedded components as
the set of polynomials vanishing to degree $s_l \kappa$ on the
varieties $V(P^\sharp_l)$ in $R^\sharp_n$, where the relation between
$Q_l$ and $P_l^\sharp$ is given by
Eq.~\eqref{eq:primary-as-power-of-prime-in-extension}.

Let us now consider applications to the problem at hand. Starting from
the first non-radical ideal that we encountered in
Eqs.~\eqref{eq:codim2-two-particle-invariant-with-longer-spinor-string}
and \eqref{eq:non-radical-ideal-oneseven}, for the ``sharp'' prime ideal in the
ring extension we can write,
\beq\label{eq:sharp-soft-ideal}
\big\langle \sqrt{\langle 17 \rangle}, |7\rangle \big\rangle_{R^\sharp_7} \, ,
\eeq
such that taking the second power we obtain,
\beq
\big\langle \sqrt{\langle 17 \rangle}, |7\rangle \big\rangle_{R^\sharp_7}^2 = \big\langle \langle 17 \rangle, |7\rangle \sqrt{\langle 17 \rangle}, |7\rangle\langle7| \big\rangle_{R^\sharp_7} \, .
\eeq
Then, intersecting with $R_n$, we have,
\beq\label{eq:first-explicit-example-of-P-sharp}
\big\langle \sqrt{\langle 17 \rangle}, |7\rangle \big\rangle_{R^\sharp_7}^{2} \cap R_7 = \big\langle \langle 17 \rangle, |7\rangle\langle7| \big\rangle_{R_7} \, ,
\eeq
that is, we have effectively removed the generator involving the
radical. This is an explicit example of the form of
Eq.~\eqref{eq:primary-as-power-of-prime-in-extension}. From this
construction we observe that the phase-space point required to infer
membership to symbolic powers of the ideal of
Eq.~\eqref{eq:sharp-soft-ideal} is such that,
\beq
\sqrt{\langle 17 \rangle} \sim \epsilon \, , \; |7\rangle \sim \epsilon \, \Rightarrow \, \langle 17 \rangle \sim \epsilon^2\, , \; |7\rangle \sim \epsilon \, ,
\eeq
while a standard ``symmetric'' approach to $V\left(\big\langle |7\rangle  \big\rangle\right)$ reads,
\beq
\langle 17 \rangle \sim \epsilon \, , \; |7\rangle \sim \epsilon \, .
\eeq
An analogous construction can be followed for the ideals in
Eqs.~\eqref{eq:non-radical-delta} and \eqref{eq:radical-delta}, for
example we can write,
\beq\label{eq:sharp-radical-delta}
\big\langle \sqrt{\langle 12 \rangle}, (s_{567}-s_{34}) \big\rangle_{R^\sharp_7} \, ,
\eeq
where we stress that the different meaning of $R^\sharp_7$ between
Eq.~\eqref{eq:sharp-radical-delta} and
Eq.~\eqref{eq:sharp-soft-ideal}. In this case, we also have that the
ideal is of maximal codimension, thus its symbolic powers must
coincide with normal powers. Once again, we have constructed the
desired ideal,
\beq
\big\langle \sqrt{\langle 12 \rangle}, (s_{567}-s_{34}) \big\rangle_{R^\sharp_7}^{ 2 } \cap R_7 = \big\langle \langle 12 \rangle, \Delta_3(1, 2, 3, 4) \big\rangle_{R_7} \, .
\eeq

Finally, as promised, the same approach can also be employed
independently of whether an ideal is radical or not, in order to
obtain futher data regarding the pole structure of the integral
coefficients. For instance, let us reconsider the primary
decomposition of Eq.~\eqref{eq:primary_decomposition_zaa22_zbb22}.
Within a suitable ring extension $R^\sharp_7$, we can write,
\begin{align}
\big\langle ⟨7|\Gamma_{34|56}|7⟩, \sqrt{[7|\Gamma_{34|56}|7]} \big\rangle_{R^\sharp_7} &= \big\langle ⟨7|\Gamma_{34}|7], ⟨7|\Gamma_{56}|7], ⟨7|\Gamma_{34|56}|7⟩, \sqrt{[7|\Gamma_{34|56}|7]} \big\rangle_{R^\sharp_7} \quad \nonumber \\
&\phantom{=} \quad \cap  \; \big\langle  ⟨7|\Gamma_{34|56}|7⟩, \sqrt{[7|\Gamma_{34|56}|7]}, \tilde\Gamma_{12|34|56} \big\rangle_{R^\sharp_7} \, ,
\end{align}
where we verified again in $R^\sharp_7$ the equality as well as the
primality of the ideals in the RHS. We can than combine constraints
from the simultaneous membership to a symbolic power of $\big\langle
⟨7|\Gamma_{34|56}|7⟩, [7|\Gamma_{34|56}|7] \big\rangle_{R_7}$
and $\big\langle ⟨7|\Gamma_{34|56}|7⟩, \sqrt{[7|\Gamma_{34|56}|7]}
\big\rangle_{R^\sharp_7}$ to obtain refined partial fraction
decompositions. To see this in practice, let us refer back to section~\ref{Integral_Coefficients}.
It can be seen that a number of
coefficients, such as $d^{(2,A)}_{\{12\times34\times56\}}$ in Eq.~\eqref{d123456m2asy}, $\tilde
d^{(2)}_{\{12\times34\times56\}}$ in Eq.~\eqref{dt123456m2} and $\tilde c^{(2)}_{\{7\times12\}}$ in Eq.~\eqref{ct712m2}, have a double
pole on $V\left(\left\langle ⟨7|\Gamma_{34|56}|7⟩
\right\rangle\right)$ and a simple pole on $V\left(\left\langle
                   [7|\Gamma_{34|56}|7] \right\rangle\right)$, i.e.~they read,
\beq
\mathcal{C}_i \propto \frac{\mathcal{N}_i}{⟨7|\Gamma_{34|56}|7⟩^2[7|\Gamma_{34|56}|7]} \, .
\eeq
Their numerator in least common denominator form belongs
to $\big\langle ⟨7|\Gamma_{34|56}|7⟩, [7|\Gamma_{34|56}|7]
\big\rangle_{R_7}$, therefore we can write them as,
\beq
\mathcal{C}_i \propto \frac{\mathcal{N}_{i1}}{⟨7|\Gamma_{34|56}|7⟩^2} + \frac{\mathcal{N}_{i2}}{⟨7|\Gamma_{34|56}|7⟩[7|\Gamma_{34|56}|7]} \, .
\eeq
However, by probing them in the asymmetric approach we also obtain the
constraint,
\beq
\mathcal{N}_i \in \big\langle \, ⟨7|\Gamma_{34|56}|7⟩, \sqrt{[7|\Gamma_{34|56}|7]} \, \big\rangle^{\langle 2 \rangle}_{R^\sharp_7} \cap R_7 \, .
\eeq
As this is a maximal codimension ideal the symbolic power coincides
with the normal power. Thus, we conclude that a more accurate
representation is,
\beq
\mathcal{C}_i \propto \frac{\mathcal{N}_{i1}}{⟨7|\Gamma_{34|56}|7⟩^2} + \frac{\mathcal{N}_{i2}}{[7|\Gamma_{34|56}|7]} \, .
\eeq

\bibliography{ms}
\bibliographystyle{JHEP}
\end{document}